\documentclass{article}

\usepackage{amsmath,amsfonts,bm}

\def\eqref#1{equation~\ref{#1}}

\def\1{\bm{1}}

\DeclareMathAlphabet{\mathsfit}{\encodingdefault}{\sfdefault}{m}{sl}
\SetMathAlphabet{\mathsfit}{bold}{\encodingdefault}{\sfdefault}{bx}{n}

\newcommand{\R}{\mathbb{R}}

\usepackage{arxiv}

\usepackage[utf8]{inputenc} 
\usepackage[T1]{fontenc}    
\usepackage{hyperref}       
\usepackage{url}            
\usepackage{booktabs}       
\usepackage{amsfonts}       
\usepackage{nicefrac}       
\usepackage{microtype}      
\usepackage{cleveref}       
\usepackage{graphicx}
\usepackage{natbib}
\usepackage{doi}

\usepackage{amssymb}
\usepackage{amsmath}
\usepackage{multirow}
\usepackage{makecell}
\usepackage{xcolor}
\usepackage{enumitem}
\usepackage{soul}
\usepackage{algorithm}
\usepackage{algpseudocode}

\newcommand{\ours}{\mbox{$\varepsilon$ar-VAE2}}
\newcommand{\earvae}{\mbox{$\varepsilon$ar-VAE}}
\newcommand{\specsnake}{Spec-SnakeBeta}
\newcommand{\refiner}{Duplex-Aware Refiner}
\DeclareRobustCommand{\unconstrainedrefiner}{Unconstrained Refiner}
\DeclareRobustCommand{\stftdist}{STFT Distance}
\DeclareRobustCommand{\logstftdist}{STFT$_{\log}$ Distance}
\DeclareRobustCommand{\meldist}{Mel Distance}
\DeclareRobustCommand{\logmeldist}{Mel$_{\log}$ Distance}
\DeclareRobustCommand{\stftdisttab}{STFT}
\DeclareRobustCommand{\logstftdisttab}{STFT$_{\log}$}
\DeclareRobustCommand{\meldisttab}{Mel}
\DeclareRobustCommand{\logmeldisttab}{Mel$_{\log}$}
\DeclareRobustCommand{\hfdisttab}{STFT$_{HF}$}
\DeclareRobustCommand{\lfdisttab}{STFT$_{LF}$}
\newcommand{\distcol}[1]{#1$\downarrow$}

\providecommand{\C}{\mathbb{C}}

\title{Fourier is Frontier: \\ Frequency-Aware Autoencoding \\ for High-Fidelity Music Reconstruction}

\date{}

\author{ Kangdi Wang \\
	Qwen Team, Alibaba \\
	\texttt{wangkangdi21@gmail.com} \\
	\And
	Yusheng Dai \\
	Monash University \\
	\texttt{yusheng.dai@monash.edu} \\
	\And
	Jin Xu\thanks{Corresponding author.} \\
	Qwen Team, Alibaba \\
	\texttt{jxu3425@gmail.com} \\
}

\renewcommand{\undertitle}{}
\renewcommand{\shorttitle}{\ours{}}

\begin{document}

\maketitle

\begin{abstract}
Continuous-latent audio autoencoders form the backbone of latent music generators, and
their reconstruction fidelity limits the acoustic detail available to downstream generators. At the
compression rates these pipelines require, decoders commonly exhibit three recurring failure
modes: high-frequency loss, phase incoherence, and stereo-image collapse. 
These shortcomings in high-fidelity generation share a structural root: waveform autoencoders, previously regarded as the frontier design, lack an explicit frequency axis, leaving no handle for targeted per-band correction. Among five matched-budget
representations in our controlled study, the complex STFT achieves the lowest full-band and high-frequency spectral
distances, providing direct access to magnitude and phase at every frequency bin.
Building on this finding, we present \ours{}, a complex-spectral autoencoder that exploits the explicit frequency structure and introduces cross-channel interaction between the left and right channels. First, \specsnake{} exploits the additional capacity of the frequency-domain representation by learning a periodic activation for each frequency bin, with frequency-dependent initialization in log-parameter space. This design overall outperforms the other activation variants in our controlled ablation,
while using fewer parameters than the fully independent variant. Second, \refiner{} applies band-specific corrections to magnitude and phase. Its band allocation
follows duplex theory of sound localization and corrects the decoder's reconstruction error.
On the 546-track Song Describer Dataset, \ours{} achieves the best point estimates on five of seven reconstruction metrics and matches the best stereo-coherence score.
Adding the \refiner{} module further reduces \meldist{} by 19.4\%, and its duplex band allocation uses ${\approx}45\%$ fewer
residual-output dimensions than the \unconstrainedrefiner{} version. Beyond this efficiency, the banded allocation also lowers spectral distances and targeted spatial-cue error metrics, and it receives higher mean paired ratings from
professional mixing and mastering engineers. The downstream generator using \ours{} latents achieves better point estimates on all 12 automatic downstream metrics.
Demo page is available at \url{https://eps-acoustic-revolution-lab.github.io/EAR_VAE2/}.
\end{abstract}

\section{Introduction}
\label{sec:intro}

Latent-diffusion pipelines are widely used for high-quality music
generation~\citep{rombach2022stablediffusion,evans2024stableaudioopen}. In these
systems, a diffusion model operates not on raw audio but on the continuous latent
space of a variational autoencoder (VAE), and a decoder maps sampled latents back
to a waveform. The decoder therefore limits the acoustic detail available to the
downstream system. Indeed, increasing the generator's capacity cannot restore spectral,
phase, or stereo information that the autoencoder does not preserve. Under aggressive
compression, decoders often lose high-frequency content, degrade phase coherence, and
narrow the stereo image. These errors are especially relevant to 48\,kHz stereo music,
which contains dense harmonics, sharp transients, and spatial cues. Waveform
VAEs~\citep{evans2024stableaudioopen,wang2025earVAE} do not expose a fixed frequency
axis for direct per-band parameterization. Many earlier spectral autoencoders instead
discarded phase and used a separate vocoder for synthesis.
This leads to our central question: \emph{which modeling choices best preserve
perceptually salient information at a fixed compression rate?}

To probe this question empirically, we compare five
audio representations under matched latent shape and rate, parameter budget, data, and
optimization (\S\ref{sec:input_repr_exp}). Complex STFT achieves the lowest full-band
and high-frequency spectral distances, whereas waveform patches perform better on
waveform alignment and low-frequency reconstruction. The comparison therefore
identifies wideband spectral detail as the strength of the spectral representation,
and this advantage, together with the physical frequency axis it exposes, motivates
our spectral-domain design.

Building on this result, we present \ours{}, a spectral-domain music autoencoder
that preserves the energy distribution of stereo music across both the stereo field and the frequency spectrum, while markedly reducing the metallic high-frequency artifacts common in prior designs.
It targets the three failure modes through explicit
stereo separation in the complex spectrum, a frequency-aware nonlinearity, and a
psychoacoustic refiner. \ours{} compresses 48\,kHz stereo music
into a 25\,Hz sequence of 128-dimensional continuous latents ($1920\times$ temporal
downsampling). Its two central contributions are frequency-aware components that the
spectral formulation enables.

We make the network's nonlinearity frequency-aware. \specsnake{}
(\S\ref{sec:specsnakebeta}) is designed to improve frequency-dependent modeling.
BigVGAN applies the Snake activation with periodic-residual parameters learned per
channel~\citep{lee2023bigvgan}; the commonly used SnakeBeta variant retains this
channel-wise parameterization. \specsnake{}
(\S\ref{sec:specsnakebeta}) instead learns one parameter pair per bin, exploiting
the exposed frequency axis. This frequency-specific, channel-shared form overall outperforms both channel-wise SnakeBeta and a fully independent channel--frequency parameterization in the controlled ablation.

The STFT offers a second advantage: it exposes a physical frequency axis while the
model compresses along time. Each bin retains a fixed physical interpretation, so a
refiner can apply different corrections to different frequency bands. Duplex theory
provides a prior for this design: low-frequency localization relies mainly on
interaural timing and phase cues, whereas high-frequency localization relies mainly
on interaural level cues. Accordingly, the \refiner{} (\S\ref{sec:refiner}) applies phase-only
correction below $1.5$\,kHz, joint magnitude-and-phase correction between $1.5$ and
$4$\,kHz, and magnitude-only correction above $4$\,kHz. It uses fewer output residual
dimensions than the \unconstrainedrefiner{}.

The full system achieves the best point estimates on five of seven reconstruction metrics on the Song Describer Dataset~\citep{manco2023thesong} and ties for the best stereo coherence. The \refiner{} reduces \meldist{} by 19.4\%. In downstream generation,
replacing the LeVo~2 VAE with \ours{} yields higher point estimates on all reported
automatic metrics under the matched evaluation protocol.

Our contributions are:
\begin{enumerate}[leftmargin=*,itemsep=2pt,topsep=2pt]
  \item A \textbf{frequency-aware continuous audio VAE model} operating on the
        complex STFT\@. In a matched-budget study, this representation gives the
        lowest full-band and high-frequency spectral distances
        (\S\ref{sec:input_repr_exp}).
  \item \textbf{\specsnake{}}, a per-bin periodic activation shared across feature
        channels with frequency-proportional log-space initialization. Across three
        seeds, it outperforms its fully independent counterpart on all six ablation
        metrics with ${\approx}128\times$ fewer activation parameters (\S\ref{sec:specsnakebeta},
        \S\ref{sec:ablation_activation}).
  \item The \textbf{\refiner{}}, a duplex-guided residual module that improves
        spectral and spatial fidelity with ${\approx}45\%$ fewer output residual dimensions
        than the \unconstrainedrefiner{} (\S\ref{sec:refiner},
        \S\ref{sec:ablation_bandmode}).
\end{enumerate}

\section{Related Work}
\label{sec:related}

\subsection{Discrete neural codecs}
The convolutional encoder--decoder with residual vector quantization (RVQ),
introduced by SoundStream~\citep{zeghidour2021soundstream} and refined by
EnCodec~\citep{defossez2022encodec} and DAC~\citep{kumar2023dac}, is the
architectural ancestor of most modern audio tokenizers. By quantizing intermediate
features into discrete codes at low bitrates, these models produce tokens suited
to storage, transmission, and autoregressive generation. The same
discreteness, however, fits latent diffusion poorly: quantization commits the
representation to a discrete interface, which decreases the reconstruction quality 
that downstream generation depends on.

\subsection{Continuous waveform autoencoders}
Continuous-latent autoencoders remove quantization in favor of a smooth latent.
RAVE~\citep{caillon2021rave} learns a KL-regularized continuous VAE latent. More
recently, Stable Audio Open~\citep{evans2024stableaudioopen},
Music2Latent~\citep{pasini2024music2latent}, \earvae{}~\citep{wang2025earVAE},  and LeVo~2~\citep{lei2026levo2stablemelodious}
provide the continuous representations on which modern latent-diffusion music
pipelines run. These latents align naturally with diffusion training and match or
exceed discrete codecs in downstream generative quality.

These models, however, encode raw waveforms, entangling frequency, phase, and
stereo cues sample by sample. Without an explicit frequency axis, the model does not
expose direct per-band parameterization by spectral location. This motivates our move
to the spectral domain for 48\,kHz stereo music.

\subsection{Revisiting the Spectral Domain}
In a spectral representation, each axis corresponds to a fixed physical frequency,
enabling direct per-band operations that waveform models do not natively support. 
Earlier spectral autoencoders nevertheless relied on waveform-domain synthesis because 
they operated on magnitude or mel spectrograms, discarding phase and delegating 
synthesis to a separate
vocoder~\citep{siuzdak2024vocos,kong2020hifigan,lee2023bigvgan}. In those systems,
the vocoder limited final reconstruction fidelity.

Within spectral-domain audio modeling, recent work has focused on conditional
vocoding~\citep{siuzdak2024vocos} or discrete
coding~\citep{ai2024apcodec,li2025spectrostream}, leaving spectral representations
underexplored for continuous latent autoencoding. Existing representation studies
either target GAN generation~\citep{nistal2020comparing} or change the quantizer
together with the representation~\citep{langman2024spectral}, making it difficult
to isolate the effect of the representation itself.

Our controlled study addresses this gap by comparing representation choices under
matched budgets (\S\ref{sec:input_repr}). It identifies the complex STFT as the best
fit for full-band and high-frequency spectral preservation, although waveform
representations retain advantages on other metrics. Guided by this comparison, we
adopt the \emph{complex STFT} for our continuous autoencoder. Its real and imaginary
components retain phase information and enable direct iSTFT synthesis without a
separate vocoder. The resulting physical frequency axis further supports the
frequency-aware operations developed in \S\ref{sec:method}.

\section{Method}
\label{sec:method}

\begin{figure}[tbh]
  \centering
  \includegraphics[width=\textwidth]{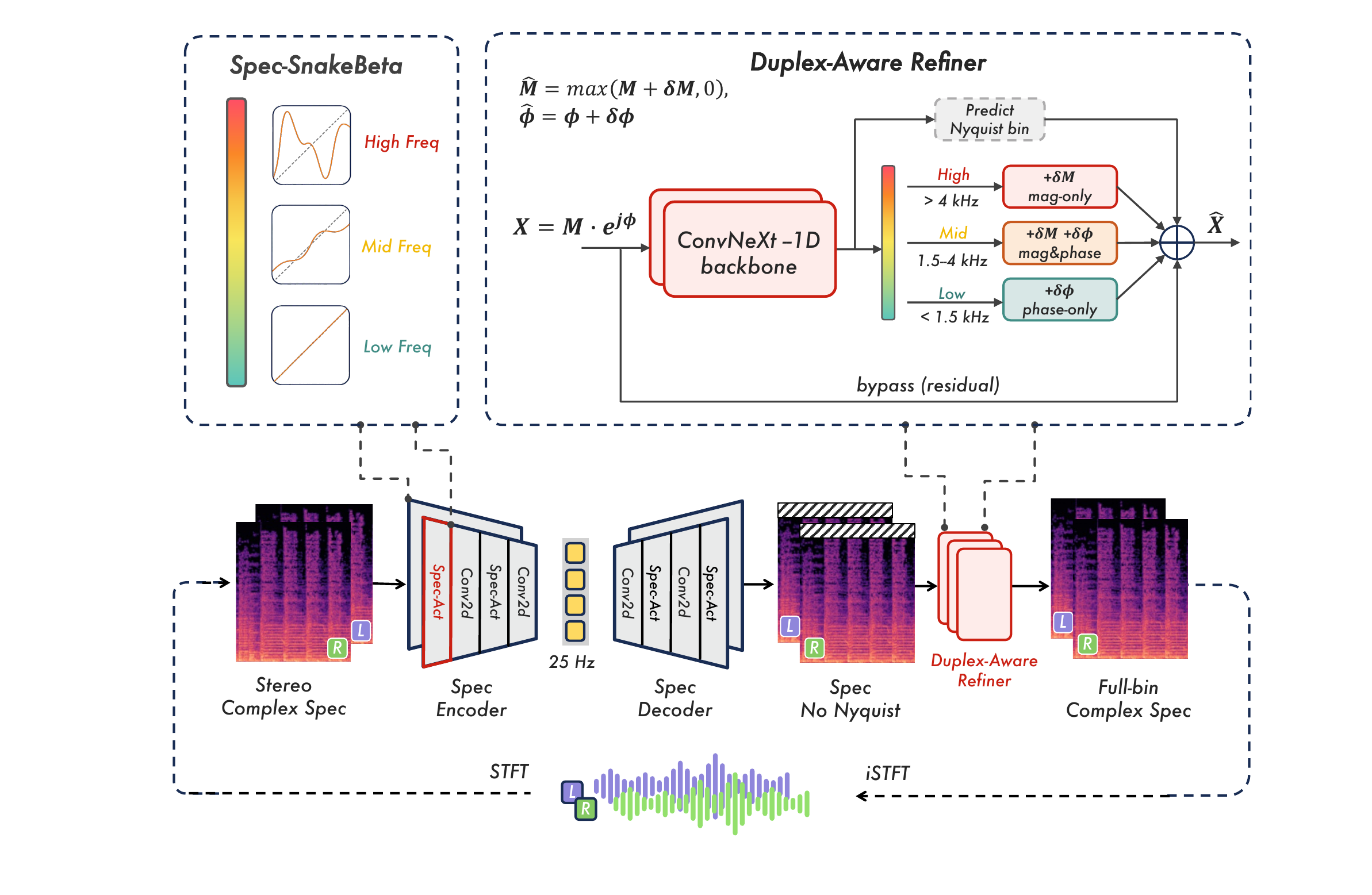}
  \caption{Architecture of \ours{}. Stereo audio is transformed via STFT into a complex
  spectrogram; a Spec Encoder compresses it to a 25\,Hz continuous latent and a Spec
  Decoder reconstructs a coarse complex spectrogram; the highest-frequency (Nyquist)
  STFT bin is removed to facilitate frequency downsampling. The \refiner{} subsequently
  applies band-specific complex corrections and restores the Nyquist bin to produce the
  full-bin spectrogram, which the iSTFT resynthesizes into a waveform.}
  \label{fig:arch}
\end{figure}

\ours{} is a spectral-domain VAE-GAN with explicit stereo modeling. 
A complex-STFT encoder--decoder produces a continuous latent and a coarse reconstruction. 
The frequency axis is exploited at two complementary levels: 
\specsnake{} adapts the internal nonlinear responses across frequency, 
while the \refiner{} applies frequency-dependent output corrections.
The refiner is trained after the encoder--decoder and operates on its frozen reconstruction.
Figure~\ref{fig:arch} provides an overview.

\subsection{Spectral Encoder--Decoder}
\label{sec:architecture}

We transform 48\,kHz stereo audio into a complex spectrogram using an STFT with
$n_\text{fft}{=}960$ ($\Delta f{=}50$\,Hz), $\text{hop}{=}480$, and a Hann window.
The one-sided STFT of each real-valued channel contains $F_\mathrm{full}{=}481$ bins; we exclude the
Nyquist bin and operate on the remaining $F{=}480$ retained non-Nyquist frequency bins.
The result is $\mathbf{X}\in\C^{2\times F\times T}$, where the leading dimension indexes
the left and right stereo signals, $f$ indexes frequency bins, $t$ indexes temporal
frames, and $T$ is the number of STFT frames at 100\,Hz. Here, $F{=}480$ denotes the
input frequency size. Intermediate layers downsample the frequency axis; $F_\ell$
denotes the number of frequency positions at layer $\ell$. $C$ later denotes the network feature-channel count, not the two
stereo channels of $\mathbf{X}$. The encoder maps $\mathbf{X}$ to a continuous latent
$\mathbf{z}\in\R^{128\times T'}$ at 25\,Hz. The decoder reconstructs a
coarse complex spectrogram, which is converted to audio by iSTFT\@. We regularize the
latent against a unit Gaussian.

Inspired by SpectroStream~\citep{li2025spectrostream}, we adopt the 2D-convolutional encoder--decoder as the architecture, but replace the residual-VQ bottleneck with
a KL-regularized continuous latent. Delayed channel fusion in the encoder and early
channel splitting in the decoder preserve explicit stereo structure. An
asymmetric-padding STFT wrapper produces exactly $T{=}L/\text{hop}$ frames while
keeping the path differentiable (Appendix~\ref{app:stft_wrapper}).

\subsection{\specsnake{} Activation}
\label{sec:specsnakebeta}

For periodic signals, the Snake activation embeds a learnable periodic residual in an otherwise identity-like activation,
improving the representation of periodic functions beyond piecewise-linear
nonlinearities~\citep{ziyin2020snake}. Building on this, BigVGAN transfers this inductive bias to
waveform synthesis, learning one periodicity per feature channel together with
anti-aliasing filters~\citep{lee2023bigvgan}. Channel-wise SnakeBeta, a decoupled
variant of this activation, $x+\beta^{-1}\sin^2(\alpha x)$, retains the per-channel
parameterization and learns one periodic response per channel. Such channel-wise
variation suits waveform features, but in a spectral network it applies the same
response across all $F$ frequency bins despite their different physical meanings. A
feature channel contains activations at many spectral locations, but channel-wise
parameters cannot vary by frequency. To our knowledge, prior work has not
parameterized this periodic prior along the spectral-frequency axis.

\begin{figure}[t]
  \centering
  \includegraphics[width=0.9\textwidth]{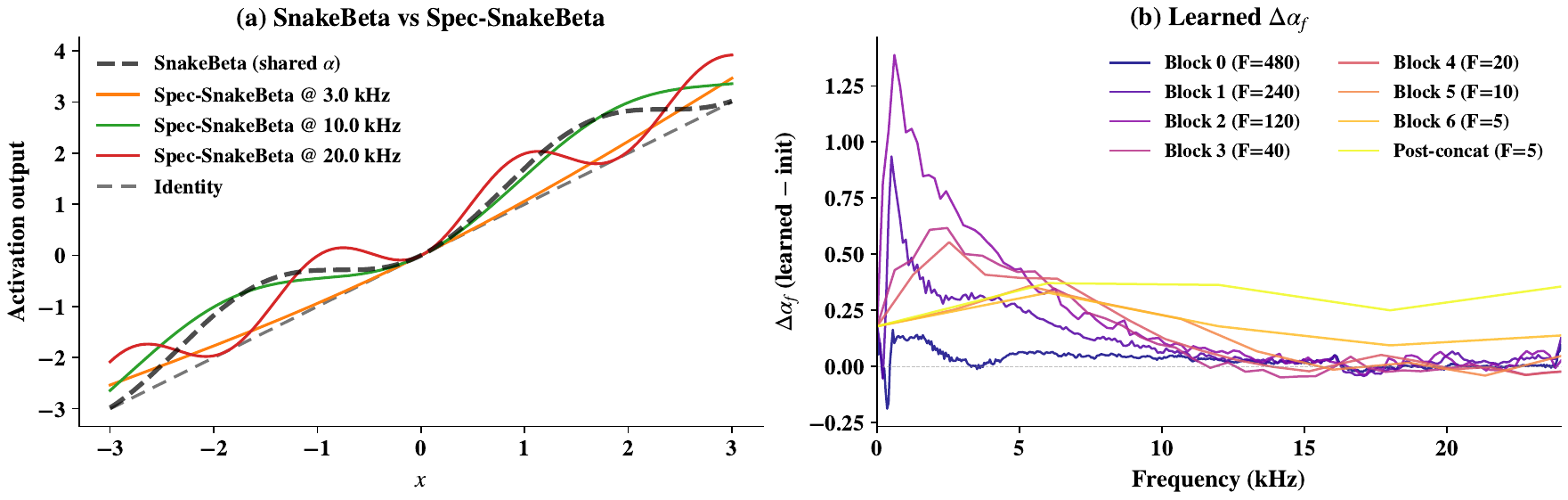}
  \caption{\specsnake{} analysis. \textbf{(a)}~Activation curves at representative
  frequencies (0.5, 3, 10, 20\,kHz) vs.\ channel-wise SnakeBeta (black dashed).
  \textbf{(b)}~Learned deviation
  $\Delta\alpha_f=\alpha^{\mathrm{trained}}_f-\alpha^{(0)}_f$
  from frequency-proportional initialization across encoder layers.}
  \label{fig:snakebeta}
\end{figure}

\specsnake{} instead learns one parameter pair per frequency position and shares it across
feature channels. Here $c$ indexes network feature channels, not the left and right
audio channels. At layer $\ell$, let $x_{b,c,t,f}$ denote the pre-activation at batch
index $b$, feature channel $c$, frame $t$, and frequency position
$f\in\{0,\ldots,F_\ell-1\}$. The frequency coordinate is downsampled with the feature
map, and $\mathrm{freq}_{\ell,f}$ denotes the corresponding center frequency. In log-space,
\begin{equation}
  \specsnake(x)_{b,c,t,f}
    = x_{b,c,t,f} + \frac{1}{\exp(\tilde\beta_f)+\epsilon_\text{num}}\,
      \sin^2\!\big(\exp(\tilde\alpha_f)\cdot x_{b,c,t,f}\big),
  \label{eq:snakebeta}
\end{equation}
with $\tilde\alpha,\tilde\beta\in\R^{F_\ell}$ for each layer and
$\epsilon_\text{num}{=}10^{-9}$. We optimize $\tilde\alpha$ and $\tilde\beta$ in log space,
with $\alpha=e^{\tilde\alpha}$ and $\beta=e^{\tilde\beta}$, so that $\alpha,\beta>0$;
$\epsilon_\text{num}$ is a numerical stabilizer in the implemented denominator, distinct
from the initialization floor $\epsilon_\text{init}$ below. We initialize
$\tilde\alpha$ proportionally to normalized frequency and $\tilde\beta$ uniformly:
\begin{equation}
  \tilde\alpha_f = \log\!\Big(\frac{\text{freq}_{\ell,f}}{\bar f_\ell}+\epsilon_\text{init}\Big),
  \qquad \bar f_\ell = \frac{1}{F_\ell}\sum_f \text{freq}_{\ell,f},
  \qquad \tilde\beta_f=0,
  \label{eq:alphainit}
\end{equation}
($\epsilon_\text{init}{=}10^{-6}$; the Nyquist bin is excluded at the input). In $\alpha$-space, the
initialization is linear in normalized frequency, while $\tilde\beta{=}0$ gives
$\beta{=}1$ at every frequency. This initialization is identity-like at low frequencies for bounded
pre-activations and increases the feature-space oscillation rate toward high
frequencies. Training adapts each frequency response while preserving a common
frequency-dependent prior across feature channels. Compared with channel-wise
SnakeBeta, this moves the learnable variation from the feature-channel axis to the
physical-frequency axis. Log-space parameterization gives the chain-rule rescaling
$\partial\mathcal{L}/\partial\tilde\alpha_f=\alpha_f\,
\partial\mathcal{L}/\partial\alpha_f$ and
$\partial\mathcal{L}/\partial\tilde\beta_f=\beta_f\,
\partial\mathcal{L}/\partial\beta_f$; these identities are not convergence claims. Because
$\alpha$ and $\beta$ are decoupled, the activation need not remain monotone; the
identity skip nevertheless centers the derivative at one, with its range controlled by
$\alpha/(\beta+\epsilon_\text{num})$. 
Figure~\ref{fig:snakebeta}(a) illustrates the effect: the
frequency-proportional initialization renders the activation identity-like at low
frequencies for the plotted input range and
progressively more oscillatory toward high frequencies. After training, the learned
$\alpha_f$ deviates from its initialization primarily in the
0--5\,kHz range
(Figure~\ref{fig:snakebeta}(b)). In sum, \specsnake{} uses the physical frequency axis 
inside the encoder--decoder, allowing different spectral regions to learn different periodic responses.

\subsection{\refiner{}}
\label{sec:refiner}

\begin{figure}[tbh]
  \centering
  \includegraphics[width=\textwidth]{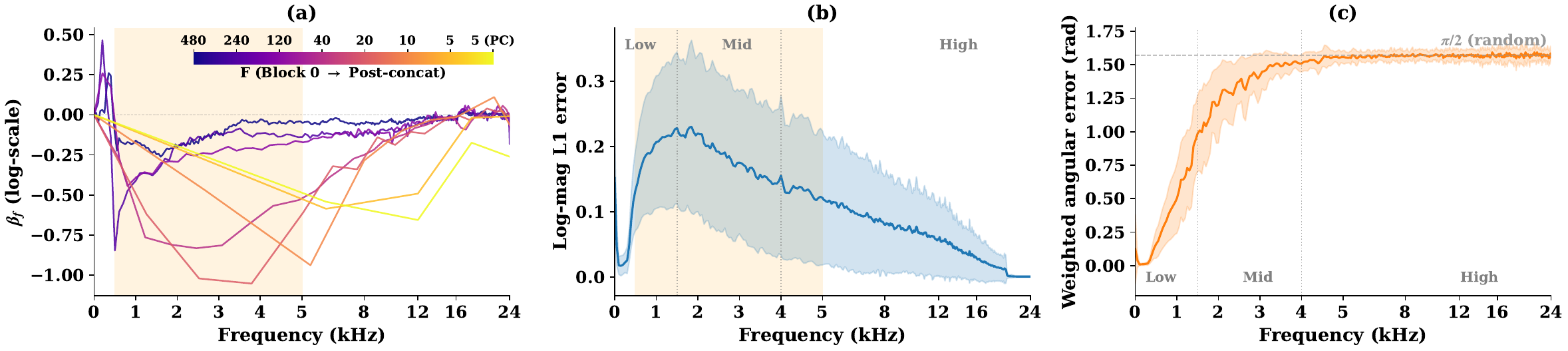}
  \caption{Frequency-resolved diagnosis of the trained encoder--decoder before
  refinement. \textbf{(a)}~Learned frequency-indexed $\tilde\beta_f$ per encoder
  layer; since $\tilde\beta_f$ is initialized to zero, the values also represent
  displacements from initialization. \textbf{(b)}~Per-frequency magnitude error.
  \textbf{(c)}~Weighted angular error; the horizontal line marks the $\pi/2$
  random-phase baseline. Vertical lines mark the refiner band boundaries, and the
  shaded region spans 0.5--5\,kHz.}
  \label{fig:error_analysis}
\end{figure}

The encoder--decoder already adapts its internal responses across frequency, yet its
frozen output retains frequency-structured reconstruction errors.
Figure~\ref{fig:error_analysis} separates these two observations. Panel~(a) shows
the learned \specsnake{} parameter $\tilde\beta_f$; because it is initialized to
zero, the plotted values also give its displacement from initialization. This
frequency-dependent adaptation occurs inside the encoder--decoder. In contrast,
panels~(b--c) measure the error that remains at its output before refinement. 
Together, these observations motivate a post-hoc refiner that corrects the
residual output errors.

Our core design insight for the refiner comes from duplex theory, which provides
both the cue allocation and its frequency boundaries: human localization relies
primarily on interaural timing and phase cues below approximately $1.5$\,kHz,
passes through a transition region from $1.5$ to $4$\,kHz, and relies increasingly
on interaural level cues above $4$\,kHz
\citep{rayleigh1907duplex,macpherson2002duplex}. Interestingly, we find that the
frozen decoder's error profile independently follows these theoretically defined
regions: magnitude error is largest in the transition band, while phase error
rises toward its plateau near the upper boundary. This convergence of theoretical
prior and empirical measurement motivates phase-only correction in the Low band,
joint magnitude-and-phase correction in the Mid band, and magnitude-only
correction in the High band.\footnote{These refiner bands (Low/Mid/High,
boundaries at $1.5$ and $4$\,kHz) serve a different purpose than the HF
($>8$\,kHz) / LF ($<2$\,kHz) metric bands defined in \S\ref{sec:metrics}: the
former guide correction placement, the latter isolate evaluation regions.}

As shown in Figure~\ref{fig:arch}, the refiner processes the decoder's complex STFT
$X=M e^{j\phi}$ with a lightweight ConvNeXt-1D backbone~\citep{liu2022convnext}.
The same module is applied independently to each stereo channel. Its shared output
projection allocates residuals by frequency band: the Low band ($<1.5$\,kHz) receives
$\delta\phi$, the Mid band ($1.5$--$4$\,kHz) receives $\delta M$ and $\delta\phi$,
and the High band ($>4$\,kHz) receives $\delta M$. The projection also predicts the
excluded Nyquist bin.

For each stereo channel, the batched input is $X\in\C^{B\times F\times T}$. The
backbone maps its real and imaginary components to per-frame residuals over the
retained frequency bins. Suppressing the batch, time-frame, and frequency-bin
indices, each retained bin is updated as
\begin{equation}
  \hat M=\max(M+\delta M,\epsilon),
  \qquad
  \hat\phi=\phi+\delta\phi,
  \qquad
  \hat X=\hat M e^{j\hat\phi}.
\end{equation}
Here, $\epsilon>0$ ensures numerical stability, $\delta M{=}0$ in the Low band, and
$\delta\phi{=}0$ in the High band. The predicted Nyquist bin is appended to these
updated bins to form the corrected spectrum. This banded allocation uses fewer output
residual dimensions than the \unconstrainedrefiner{}; exact dimensions appear in
Appendix~\ref{app:arch}.

All refiner output weights are zero-initialized, so $\hat X{=}X$ when training begins
and adding the refiner preserves the trained decoder output. The encoder and decoder are frozen, and the refiner learns only the additional band-specific
correction.

\subsection{Training Objective and Discriminators}
\label{sec:losses}

\subsubsection{Losses}
The objective has four base terms. A multi-resolution STFT loss~\citep{yamamoto2020parallel}
combines spectral convergence and log-magnitude L1 at seven FFT scales from 32 to
2048. An Instantaneous Frequency/Group Delay (IF/GD) phase loss follows
\earvae{}~\citep{wang2025earVAE} and computes cosine distance on conjugate products
along the time and frequency axes. The adversarial (Adv) term combines an LSGAN
loss~\citep{mao2017lsgan} with feature matching (FM)~\citep{larsen2016vaegan}.
The remaining term is KL regularization against a
unit Gaussian. Beyond these four terms, we adopt several perceptual techniques:
K-weighting~\citep{itu2015bs1770}, Mid/Side/Left/Right (MSLR) stereo
decomposition~\citep{wang2025earVAE}, the adaptive log-magnitude normalization of
SAME~\citep{parker2026samesemanticallyalignedmusicautoencoder}, and the mixed-scale spectral loss of
SpectroStream~\citep{li2025spectrostream}. All losses use float64 precision to improve numerical stability in complex-domain calculations.

\subsubsection{Discriminators}
We use three discriminator families along two supervision paths. The waveform-routed
path contains multi-scale STFT and constant-Q transform (CQT) discriminators, which operate on the waveform
after iSTFT and backpropagate through the synthesis path; we denote them collectively
as $D_w$. The direct spectral path uses a spectral discriminator $D_s$ on the
refiner's native complex STFT output before resynthesis.

The \textbf{multi-scale STFT discriminator} addresses the inability of a single STFT
resolution to capture both short transients and fine harmonic structure. Five
sub-discriminators use dyadic FFT sizes
$n_\text{fft}\in\{128,256,512,1024,2048\}$ with
$\text{hop}{=}n_\text{fft}/2$, maintaining 50\% overlap while spanning multiple
time--frequency scales. Three additional sub-discriminators fix
$n_\text{fft}{=}4096$ and use hops $\{2048,1024,256\}$. At 48\,kHz, this FFT size
gives $\Delta f{\approx}11.7$\,Hz; varying the hop provides three temporal sampling
densities at the same frequency resolution. For the per-channel waveform $w$, each
sub-discriminator receives
$[\operatorname{Re}(S_r(w)),\operatorname{Im}(S_r(w)),|S_r(w)|]$, where $S_r$
denotes its STFT configuration. We retain this dyadic FFT grid instead of the
golden-ratio spacing used by StableCodec~\citep{parker2025stablecodec}. The dyadic
design preserves a regular multi-resolution schedule and permits efficient
power-of-two FFTs; in our setting, golden-ratio spacing incurred additional
computation without yielding consistent metric improvements.

The \textbf{CQT discriminator} complements these linear-frequency views with a
log-frequency representation aligned with musical pitch. It contains three
sub-discriminators with $\{24,36,48\}$ bins per octave over nine octaves.

The \textbf{spectral discriminator} $D_s$, introduced only in Stage~3, evaluates the
refiner's native complex STFT output using adaptive frequency pooling at four scales.
Waveform-routed objectives can constrain only changes observable after iSTFT, whereas
$D_s$ directly regularizes native spectral directions that are not identifiable
through the waveform path. At the Stage~3 boundary, the identity-initialized refiner
reproduces the frozen Stage~2 decoder output, on which $D_w$ has already been trained.
After its warmup, $D_s$ therefore provides a newly learned, complementary
direct-spectral signal for optimizing the residual correction. Appendix~\ref{app:spec_disc}
analyzes these two motivations through synthesis-null-space and Stage~3 boundary
probes.

\subsection{Three-Stage Training Pipeline}
\label{sec:training}

Training has three stages: reconstruction pretraining; waveform-adversarial training
with STFT and CQT discriminators; and refiner-only training with STFT and IF/GD losses,
the retained $D_w$, and the added $D_s$. Stage~1 establishes the KL-regularized reconstruction before adversarial gradients are
introduced. Stage~2 then optimizes the full encoder--decoder under $D_w$.
Finally, Stage~3 freezes that backbone, drops KL, retains $D_w$, adds $D_s$, and
trains only the refiner with phase, reconstruction, and adversarial supervision.
This separation prevents the refiner from destabilizing the learned latent space.
For optimizers, stages~1--2 use the convolution-adapted Muon optimizer~\citep{jordan2024muon} with AdamW for remaining parameters; Stage~3 uses AdamW.
Schedules, update ratios, and optimizer details appear in
Appendices~\ref{app:training_config} and~\ref{app:muon}.

\section{Experiments}
\label{sec:experiments}

\subsection{Setup}
\label{sec:setup}

We evaluate reconstruction on 546 tracks in the Song Describer
Dataset~\citep{manco2023thesong}. The baselines are four recent open-source audio VAEs:
\earvae{}~\citep{wang2025earVAE}, Stable Audio Open
(SA-Open)~\citep{evans2024stableaudioopen},
LeVo~2~\citep{lei2026levo2stablemelodious}, and
SAME-L~\citep{parker2026samesemanticallyalignedmusicautoencoder}. Each system uses its
native configuration. We report \ours{} both without and with the \refiner{} to
measure the refiner's contribution. By contrast, comparisons within \ours{}, including the ablations, are
controlled. Before computing metrics, we resample each evaluation track to the native
sample rate of the model.

The full-scale \ours{} is trained on approximately two million publicly sourced tracks,
including a subset of LAION-DISCO-12M~\citep{laion2024disco12m,lanzendorfer2023disco10m}, plus 10{,}000 hours of proprietary in-house 48\,kHz music. An automated
quality pipeline filters the training data (Appendix~\ref{app:data}).

\subsection{Metrics}
\label{sec:metrics}

We report Scale-Invariant Signal-to-Noise Ratio (SI-SDR)~\citep{leroux2019sisdr}, four
spectral distances (\stftdist{}, \logstftdist{}, \meldist{}, and \logmeldist{}), and two
stereo metrics: CCPC (Cross-Channel Phase Coherence)~\citep{wang2025earVAE} and
Spectral Pan Error (SPE)~\citep{avendano2002stereo}, which averages the absolute error
in intensity-derived pan position. Higher is better for SI-SDR, CCPC, and subjective
paired ratings ($\uparrow$); lower is better for the distances and SPE ($\downarrow$).
\stftdist{} and \meldist{} are the auraloss STFT magnitude distances on linear STFT and mel magnitudes~\citep{steinmetz2020auraloss};
\logstftdist{} and \logmeldist{} apply $\operatorname{log1p}$ to those spectrograms before the same L1 distance.
These magnitude distances are distinct from spectral convergence, which auraloss implements as a separate term.
Tables abbreviate them as \stftdisttab{}, \logstftdisttab{}, \meldisttab{}, and \logmeldisttab{}.
In the input-representation study, \hfdisttab{} and \lfdisttab{} are \stftdist{} restricted to bins above $8$\,kHz (HF) and below $2$\,kHz (LF).
These HF/LF metric bands are distinct from the refiner's Low ($<$1.5\,kHz), Mid (1.5--4\,kHz), and High ($>$4\,kHz) correction bands.
Appendix~\ref{app:metrics} gives the definitions and scale settings for the metrics listed above.
The refiner ablation (\S\ref{sec:ablation_bandmode}) additionally reports band-restricted
interaural phase and level difference errors, defined in Appendix~\ref{app:duplex_cues}.

\subsection{Controlled Input-Representation Study}
\label{sec:input_repr_exp}
\label{sec:input_repr}

We first test the representation choice that motivates the model. The study compares
five compact autoencoders: complex STFT, waveform patch, waveform Conv1D,
magnitude-only, and a patch Transformer. Each model has approximately 3.0--3.1M
parameters, matched within $\pm5\%$, and uses a 128-dimensional latent at 25\,Hz. All
variants train for 100k steps with the same data, losses, optimizer, and four-scale STFT
discriminator. The complex-STFT and waveform-patch variants also share the same Conv2D
backbone. This pair directly compares an axis organized by physical frequency with one
organized by within-patch sample position. The full five-way study compares practical
representation paradigms under the same macro-level budget; Appendix~\ref{app:input_repr}
gives the complete protocol.

\begin{table}[tbh]
  \centering
  \small
  \caption{Targeted spectral results from the matched-budget representation study on
  the Song Describer Dataset. \hfdisttab{} / \lfdisttab{} are \stftdist{} above $8$\,kHz and below $2$\,kHz (metric bands, not the refiner Low/Mid/High bands). Complex STFT
  gives the lowest full-band and high-frequency distances, while waveform patch gives
  the lowest low-frequency distance. \textbf{Bold}: best;
  \underline{underline}: second best.}
  \label{tab:input_repr_targeted}
  \resizebox{0.6\textwidth}{!}{
  \begin{tabular}{@{}lccc@{}}
    \toprule
    Representation & \distcol{\stftdisttab} & \distcol{\hfdisttab} & \distcol{\lfdisttab} \\
    \midrule
    Waveform Patch (Conv2D)  & 1.294 & 1.513 & \textbf{0.986} \\
    Complex STFT (real/imag) & \textbf{1.122} & \textbf{1.178} & \underline{1.137} \\
    Magnitude-only           & \underline{1.228} & \underline{1.376} & 1.213 \\
    Waveform (Conv1D)        & 1.589 & 1.740 & 1.358 \\
    Patch Transformer        & 2.222 & 2.350 & 1.749 \\
    \bottomrule
  \end{tabular}}
\end{table}

Complex STFT achieves the lowest full-band \stftdist{} (1.122) and the lowest
high-frequency distance (1.178), the latter being 22\% lower than the waveform-patch result (Table~\ref{tab:input_repr_targeted}).
The trade-off is equally clear: waveform patch achieves the best values on three of the four aggregate metrics (SI-SDR, CCPC, and \meldist{}) and also gives the lowest low-frequency
distance. We therefore do not claim that complex STFT is uniformly better. Instead,
the study identifies it as the better fit when full-band and high-frequency spectral
preservation are the target. Appendix~\ref{app:input_repr} reports all aggregate
metrics.

The shared-backbone comparison supports input organization as a source of this
spectral advantage: in the complex-STFT model, neighboring rows carry
fixed frequency meaning. Moreover, this fixed frequency meaning is what enables
\specsnake{} and the \refiner{}.
Figure~\ref{fig:input_repr_spec} provides a qualitative comparison of the reconstructed
spectrograms.

\subsection{Main Reconstruction Results}
\label{sec:main_results}

\begin{table}[tbh]
  \centering
  \small
  \caption{Objective reconstruction on the Song Describer Dataset (546 tracks).
  \textbf{Bold}: best point estimate; \underline{underline}: second-best.
  \stftdisttab{} and \meldisttab{} follow the auraloss STFT magnitude L1 distance; \logstftdisttab{} and \logmeldisttab{} apply $\operatorname{log1p}$ to the spectrogram before the same L1 distance.
  The two \ours{} rows report the pipeline without and with the \refiner{}.
  Systems operate at their native configurations, so this is a competitive comparison rather than a matched-budget comparison.}
  \label{tab:main_results}
  \resizebox{\textwidth}{!}{
  \begin{tabular}{@{}l cc @{\hskip 6pt} c cc cc cc@{}}
    \toprule
    & \multicolumn{2}{c}{\textit{Config}} & \textbf{Temporal} & \multicolumn{2}{c}{\textbf{Spectral}} & \multicolumn{2}{c}{\textbf{Mel}} & \multicolumn{2}{c}{\textbf{Stereo}} \\
    \cmidrule(lr){2-3}\cmidrule(lr){4-4}\cmidrule(lr){5-6}\cmidrule(lr){7-8}\cmidrule(lr){9-10}
    System & SR & Lat.\ rate & SI-SDR$\uparrow$ & \distcol{\stftdisttab} & \distcol{\logstftdisttab} & \distcol{\meldisttab} & \distcol{\logmeldisttab} & CCPC$\uparrow$ & SPE$\downarrow$ \\
    \midrule
    \earvae{} & 44.1 & 43.1 & \underline{12.4} & \underline{0.880} & 0.079 & \underline{0.509} & 0.095 & \textbf{0.973} & \underline{0.267} \\
    SA-Open & 44.1 & 21.5 & 6.7 & 1.016 & 0.089 & 0.612 & 0.106 & 0.933 & 0.276 \\
    LeVo~2 & 48 & 25.0 & 8.1 & 0.971 & 0.086 & 0.599 & 0.103 & 0.947 & 0.273 \\
    SAME-L & 44.1 & 10.8 & \textbf{12.5} & 0.986 & 0.079 & 0.539 & 0.096 & \underline{0.970} & 0.276 \\
    \midrule
    \ours{} (without refiner) & 48 & 25.0 & 10.9 & 0.916 & \underline{0.078} & 0.572 & \underline{0.093} & 0.966 & 0.268 \\
    \textbf{\ours{} (with refiner)}  & 48 & 25.0 & 11.3 & \textbf{0.870} & \textbf{0.075} & \textbf{0.461} & \textbf{0.089} & \textbf{0.973} & \textbf{0.264} \\
    \bottomrule
  \end{tabular}}
\end{table}

Table~\ref{tab:main_results} summarizes the results. The full \ours{} achieves the
best point estimates on five of seven metrics (\stftdist{}, \logstftdist{},
\meldist{}, \logmeldist{}, and SPE) and matches the best CCPC\@. The refiner
accounts for the largest mel-fidelity gain (\meldist{} $0.572 \to 0.461$, a
$19.4\%$ reduction).

\subsection{Efficiency and Perceptual Value of Banded Refinement}
\label{sec:ablation_bandmode}

The refiner's central design choice is to allocate magnitude and phase corrections by
frequency band. In Table~\ref{tab:bandmode}, we compare the default duplex placement (Low-band
phase-only correction, Mid-band joint magnitude-and-phase correction, and High-band magnitude-only correction) against the
\unconstrainedrefiner{}, which predicts magnitude and phase corrections everywhere.

The banded head predicts 532 rather than 962 residual dimensions per channel, a
reduction of ${\approx}45\%$. Despite this reduction, it improves \meldist{}/\stftdist{} from 0.705/1.120 to
0.658/1.019 and reduces the band-restricted interaural phase difference (IPD) and
interaural level difference (ILD) errors by 2.4\% (Low IPD), 8.1\% (Low ILD), and
14.7\% (High ILD). These are the duplex cues that the band allocation targets;
Appendix~\ref{app:duplex_cues} defines both metrics.
Ten professional mixing and mastering engineers also give the
banded design a higher mean rating in the paired test (0.75 vs.\ 0.66; protocol in
Appendix~\ref{app:bandmode}).

\begin{table}[tbh]
	\centering
	\small
	\caption{Banded versus \unconstrainedrefiner{}. Both variants use the same frozen
	decoder and training protocol on the Song Describer Dataset. \meldisttab{} and \stftdisttab{} denote \meldist{} and \stftdist{}.
	Low/High IPD (interaural phase difference) and ILD (interaural level difference) are mean absolute deviations from the reference,
	computed over the Low ($<$1.5\,kHz) and High ($>$4\,kHz) refiner bands rather than the HF/LF metric bands of \S\ref{sec:metrics}.
	IPD errors are in radians and ILD errors in decibels, so magnitudes are comparable only within a column;
	Appendix~\ref{app:duplex_cues} gives both definitions.
	Rating is the mean paired rating on a 1--10 scale mapped to $[0,1]$. \textbf{Bold}: best.}
	\label{tab:bandmode}
	\resizebox{\textwidth}{!}{%
	\begin{tabular}{@{}l c @{\hskip 6pt} cc @{\hskip 6pt} cccc @{\hskip 6pt} c@{}}
	  \toprule
	  & \textbf{Efficiency} & \multicolumn{2}{c}{\textbf{Spectral}} & \multicolumn{4}{c}{\textbf{Duplex-Cue}} & \textbf{Subjective} \\
	  \cmidrule(lr){2-2}\cmidrule(lr){3-4}\cmidrule(lr){5-8}\cmidrule(lr){9-9}
	  & Out.\ dims/ch & \distcol{\meldisttab} & \distcol{\stftdisttab} & Low IPD$\downarrow$ & High IPD$\downarrow$ & Low ILD$\downarrow$ & High ILD$\downarrow$ & Rating$\uparrow$ \\
	  \midrule
	  \textbf{Banded (ours)} & \textbf{532} & \textbf{0.658} & \textbf{1.019} & \textbf{0.7963} & 1.2907 & \textbf{0.1690} & \textbf{0.9804} & \textbf{0.75} \\
	  Unconstrained Refiner          & 962 & 0.705 & 1.120 & 0.8155 & \textbf{1.2825} & 0.1840 & 1.1495 & 0.66 \\
	  \bottomrule
	\end{tabular}}
  \end{table}

\subsection{Ablation: \specsnake{} Activation}
\label{sec:ablation_activation}

We compare seven configurations to distinguish the effects of the
periodic prior, the parameter axis, the frequency-proportional initialization, and
channel sharing (Table~\ref{tab:activation}). As a reminder, $C$ here indexes network feature channels, not the left and right audio channels of $\mathbf{X}$.
The three parameterizations are $C$ (channel-indexed), $F$ (frequency-indexed and channel-shared), and $CF$ (channel--frequency independent).
The controlled setting uses a 42.6M-parameter generator ($C_0{=}32$ base width, $D{=}64$ bottleneck width) trained for
100k steps on the filtered LAION-DISCO-12M subset. All other hyperparameters are shared;
F-Uniform is the $F$ parameterization with zero initialization ($\tilde\alpha_f{=}\tilde\beta_f{=}0$); Uniform refers to this parameter initialization, not to frequency sampling.
F-Log is the $F$ parameterization stored in log space with frequency-proportional initialization.
CF-Log uses the F-Log initialization but optimizes every $(c,f)$ pair independently.
Thus,
F-Uniform versus F-Log isolates initialization, $C$ versus F-Uniform changes the indexed
axis, and F-Log versus CF-Log isolates channel sharing.
Appendix~\ref{app:activation} gives the full protocol.

\begin{table}[tbh]
  \centering
  \small
  \caption{Activation ablation. $C$ is channel-indexed (feature channel, not left/right audio), $F$ is frequency-indexed and channel-shared, and $CF$ is channel--frequency independent.
  \stftdisttab{} and \meldisttab{} follow the auraloss STFT magnitude L1 distance; \logstftdisttab{} and \logmeldisttab{} apply $\operatorname{log1p}$ before the same L1 distance.
  \textbf{Bold}: best; \underline{underline}: second best.}
  \label{tab:activation}
  \resizebox{0.85\textwidth}{!}{
  \begin{tabular}{@{}l c cc cc c@{}}
    \toprule
    & \textit{Temporal} & \multicolumn{2}{c}{\textit{Spectral}} & \multicolumn{2}{c}{\textit{Mel}} & \textit{Stereo} \\
    \cmidrule(lr){2-2}\cmidrule(lr){3-4}\cmidrule(lr){5-6}\cmidrule(lr){7-7}
    Variant & SI-SDR$\uparrow$ & \distcol{\stftdisttab} & \distcol{\logstftdisttab} & \distcol{\meldisttab} & \distcol{\logmeldisttab} & CCPC$\uparrow$ \\
    \midrule
    ELU                      & 2.38          & \textbf{1.313}    & 0.106             & \underline{1.114} & 0.124             & 0.886 \\
    SiLU                     & 1.78          & 1.334             & \underline{0.104} & 1.126             & \underline{0.123} & 0.883 \\
    GELU                     & 1.78          & 1.382             & 0.107             & 1.193             & 0.125             & 0.881 \\
    SnakeBeta-C              & 3.40          & 1.379             & 0.106             & 1.194             & 0.124             & \underline{0.897} \\
    \specsnake{}-F-Uniform   & 3.46          & 1.365             & 0.106             & 1.206             & \textbf{0.120}    & 0.896 \\
    \specsnake{}-CF-Log      & \underline{3.51} & 1.380          & 0.109             & 1.223             & 0.128             & 0.885 \\
    \specsnake{}-F-Log (ours)& \textbf{4.40} & \underline{1.315} & \textbf{0.102}    & \textbf{1.070}    & 0.125             & \textbf{0.908} \\
    \bottomrule
  \end{tabular}}
\end{table}

The proposed F-Log configuration achieves the best three-seed mean on four of six
metrics. Frequency-proportional initialization improves five metrics over F-Uniform,
including SI-SDR from 3.46 to 4.40\,dB. F-Log also outperforms CF-Log on all six
metrics while using 2,730 rather than 350,720 activation parameters, an
${\approx}128\times$ reduction. These
results support a shared frequency response as
an effective inductive bias rather than a parameter-saving compromise (full protocol
in Appendix~\ref{app:activation}).

\subsection{Downstream Generation}
\label{sec:downstream}

We keep the downstream renderer architecture fixed and retrain it separately for each
VAE latent space. The renderer generates VAE latents from the conditioning signal; the
corresponding VAE decoder then maps those latents to audio, followed by the optional
\refiner{} for \ours{}.
We compare two latent spaces: the LeVo~2 VAE and \ours{}. Among all 
baselines, LeVo~2 matches
\ours{} in both sample rate (48~kHz) and latent rate (25~Hz), so this pairing
keeps temporal resolution and output bandwidth fixed across the two VAE pipelines.
Each system generates 100 English and 100 Chinese songs. We score the outputs
with the seven-axis SongBench rubric~\citep{wu2026songbenchfinegrainedmultiaspectbenchmark}
and the five-axis SongEval aesthetics model~\citep{yao2025songeval}. SongBench scores
melody, arrangement, musicality, vocal, instrument, mixing, and structure. SongEval
scores coherence, musicality, memorability, structural clarity, and vocal naturalness.
Under matched conditioning and guidance, replacing the LeVo~2 VAE with \ours{} gives
higher point estimates on all 12 automatic metrics. Appending the \refiner{} at decode
time raises every point estimate again (Table~\ref{tab:gen}).

\begin{table}[tbh]
  \centering
  \small
  \caption{Automatic downstream generation evaluation under matched conditioning and
  classifier-free guidance (CFG; 100 English and 100 Chinese songs per system). The same renderer architecture is retrained for
  each latent space; the two \ours{} rows share the same latent and coarse decoder, and
  the \ours{}(-) variant omits the optional duplex-aware refiner. \textbf{Bold}: best;
  \underline{underline}: second best.}
  \label{tab:gen}
  \setlength{\tabcolsep}{4pt}
  \resizebox{\textwidth}{!}{
  \begin{tabular}{@{}l ccccccc c ccccc@{}}
    \toprule
    \multirow{2}{*}{System} & \multicolumn{7}{c}{\textbf{SongBench} ($\uparrow$)} & & \multicolumn{5}{c}{\textbf{SongEval} ($\uparrow$)} \\
    \cmidrule(lr){2-8}\cmidrule(lr){10-14}
     & \makecell{Mel-\\ody} & \makecell{Arran-\\gement} & \makecell{Music-\\ality} & \makecell{Vo-\\cal} & \makecell{Instru-\\ment} & \makecell{Mix-\\ing} & \makecell{Struc-\\ture} & & \makecell{Coher-\\ence} & \makecell{Music-\\ality} & \makecell{Memor-\\ability} & \makecell{Clar-\\ity} & \makecell{Natur-\\alness} \\
    \midrule
    LeVo~2 VAE       & 6.56 & 6.88 & 5.71 & 6.88 & 6.95 & 6.59 & 6.38 & & 4.31 & 4.17 & 4.24 & 4.20 & 4.10 \\
    \ours{}(-) & \underline{6.88} & \underline{7.17} & \underline{6.04} & \underline{7.30} & \underline{7.19} & \underline{7.02} & \underline{6.75} & & \underline{4.39} & \underline{4.25} & \underline{4.33} & \underline{4.29} & \underline{4.18} \\
    \textbf{\ours{}} & \textbf{7.02} & \textbf{7.24} & \textbf{6.15} & \textbf{7.39} & \textbf{7.23} & \textbf{7.11} & \textbf{6.81} & & \textbf{4.42} & \textbf{4.27} & \textbf{4.38} & \textbf{4.44} & \textbf{4.25} \\
    \bottomrule
  \end{tabular}}
\end{table}

\section{Discussion and Conclusion}
\label{sec:conclusion}
\label{sec:discussion}

We presented \ours{}, a frequency-aware complex-spectral autoencoder for 48\,kHz
stereo music. The explicit frequency axis supports two components: \specsnake{},
which learns one periodic nonlinearity per frequency bin shared across feature
channels, and the \refiner{}, which applies band-specific magnitude and phase
corrections guided by duplex theory. On the Song Describer Dataset, the complete system achieves
the best point estimates on five of seven reconstruction metrics and ties for the best
CCPC; the \refiner{} reduces \meldist{} by 19.4\% while using ${\approx}45\%$ fewer output
residual dimensions than the \unconstrainedrefiner{}.

The cross-system results compare native model configurations, and the downstream
evaluation gives higher automatic point estimates with \ours{} latents. The
representation, activation, and banded-refinement studies provide more controlled
comparisons of individual design choices. For example, the shared \specsnake{}
configuration outperforms its fully independent counterpart on all six metrics across
three seeds while using ${\approx}128\times$ fewer activation parameters. These
controlled results establish relative rankings rather than production-scale limits.
Separately, in an exploratory latent probe (Appendix~\ref{app:latent_probe}), \ours{} has the
largest temporal-frequency ``inversion'' ratio among the tested models. Its rapidly
varying latent component decodes to audio with a lower spectral centroid. This
observation is diagnostic; the main claims rest on the reconstruction,
ablation, spatial-cue, and expert-rating results.



\newpage
\bibliographystyle{unsrtnat}
\bibliography{references}

\newpage
\appendix
\section{Input-Representation Study: Full Protocol and Extended Results}
\label{app:input_repr}

\subsection{Protocol}
Each autoencoder has approximately 3.0--3.1M parameters, matched within $\pm5\%$ by
counting \texttt{sum(p.numel())}. Every model uses the same latent tensor
$(B,128,25{\cdot}s)$, with 128 dimensions at 25\,Hz. All variants train for 100k steps
with the same multi-resolution STFT loss, feature-matching loss, optimizer, and 4-scale
STFT discriminator. We omit the CQT discriminator to remove it as a confounding
factor. The spectral variants also share the same STFT parameters
($n_\text{fft}{=}960$, $\text{hop}{=}480$). Training uses a public music subset from
the collections described in Appendix~\ref{app:data}. We evaluate reconstruction on
the Song Describer Dataset~\citep{manco2023thesong} and report the mean metric value across files.

We compare five paradigms: \textbf{Complex STFT} (real/imaginary, Conv2D),
\textbf{Waveform Patch} (Conv2D), \textbf{Waveform} (Conv1D),
\textbf{Magnitude-only} (with a Vocos-style iSTFT head), and
\textbf{Patch Transformer}. The waveform-patch model divides the raw waveform into
480-sample patches, reshapes them into a 2D grid, and uses the same Conv2D backbone as
the complex-STFT model. Thus, the vertical axis represents physical frequency in one
model and within-patch sample position in the other. The waveform-patch model operates
entirely in the time domain (waveform $\to$ patch grid $\to$ network $\to$ grid
$\to$ waveform). It achieves the best values on the waveform-alignment metrics in this compact comparison,
while complex STFT achieves the lowest full-band and high-frequency spectral distances.

This controlled study is designed to establish the \emph{relative ranking} of
input-representation paradigms. The compact model scale means that absolute
reconstruction quality (e.g.,\ negative SI-SDR for spectral paradigms) is far from
production-level, and the production system described in the main paper is trained at
full scale. The paradigm-to-implementation mapping is given in
\S\ref{sec:input_repr}.

\subsection{Full results (five paradigms)}
Table~\ref{tab:input_repr_full} presents the full per-paradigm results. On aggregate
metrics, the waveform-patch (Conv2D) paradigm achieves the best SI-SDR, CCPC, and
\meldist{} values. Relative to complex STFT, it shows weaker full-band and
high-frequency spectral preservation, with STFT and HF distances of 1.294 and 1.513,
respectively. The patch transformer trails on every metric at this compact budget, but
replacing its Transformer backbone with the same Conv2D yields the waveform-patch
paradigm (the aggregate-metric leader), showing that patch tokenization alone does not
explain the deficit; backbone inductive bias and optimization may also contribute.

\begin{table}[tbh]
  \centering
  \caption{Input-representation study on the Song Describer Dataset (matched budget).
  \textbf{Bold}: best; \underline{underline}: second best. \hfdisttab{} / \lfdisttab{} are \stftdist{}
  above $8$\,kHz and below $2$\,kHz (metric bands, not the refiner Low/Mid/High bands).}
  \label{tab:input_repr_full}
  \label{tab:freqband}
  \resizebox{0.85\textwidth}{!}{
  \begin{tabular}{@{}lccccc cc@{}}
    \toprule
    & \multicolumn{4}{c}{\textit{Aggregate}} & & \multicolumn{2}{c}{\textit{Freq-Band}} \\
    \cmidrule(lr){2-5}\cmidrule(lr){7-8}
    Representation & SI-SDR$\uparrow$ & CCPC$\uparrow$ & \distcol{\meldisttab} & \distcol{\stftdisttab} & & \distcol{\hfdisttab} & \distcol{\lfdisttab} \\
    \midrule
    Waveform Patch (Conv2D)  & \textbf{$+$3.98} & \textbf{0.865} & \textbf{1.184} & 1.294                    & & 1.513                    & \textbf{0.986} \\
    Complex STFT (real/imag) & \underline{$-$10.49} & \underline{0.800} & \underline{1.523} & \textbf{1.122} & & \textbf{1.178} & \underline{1.137} \\
    Magnitude-only           & $-$42.16          & 0.756          & 1.568          & \underline{1.228} & & \underline{1.376} & 1.213          \\
    Waveform (Conv1D)        & $-$33.56          & 0.732          & 2.009          & 1.589             & & 1.740                    & 1.358          \\
    Patch Transformer        & $-$52.04          & 0.697          & 2.828          & 2.222             & & 2.350                    & 1.749          \\
    \bottomrule
  \end{tabular}}
\end{table}

\begin{figure}[h]
  \centering
  \includegraphics[width=\textwidth]{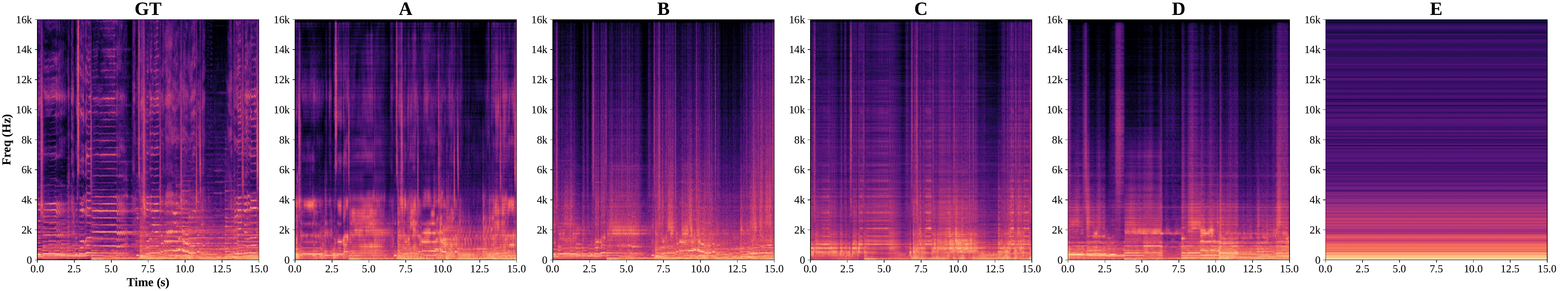}
  \caption{Reconstruction spectrograms from the matched-budget representation study
  (48\,kHz, $n_\text{fft}{=}2048$, hop${=}128$). \textbf{(A)}~complex STFT;
  \textbf{(B)}~waveform patch; \textbf{(C)}~magnitude-only;
  \textbf{(D)}~waveform Conv1D; \textbf{(E)}~patch Transformer; and
  \textbf{(GT)}~ground truth. Complex STFT and waveform patch share the same Conv2D
  backbone.}
  \label{fig:input_repr_spec}
\end{figure}

\subsection{Training stability}
\label{app:grad_norm}

\begin{figure}[h]
  \centering
  \includegraphics[width=\linewidth]{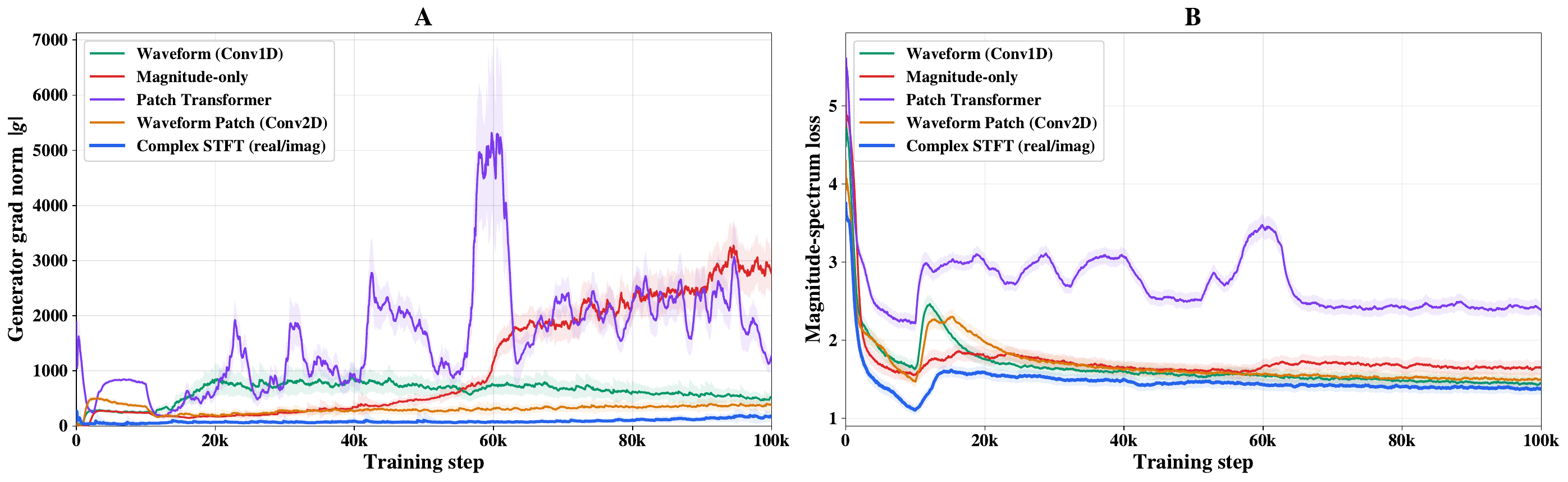}
  \caption{Encoder gradient $\ell_2$-norm (left) and magnitude reconstruction
  loss (right) over training, recorded during the same 100k-step runs under the
  shared protocol.}
  \label{fig:grad_norm_overlay}
\end{figure}

Figure~\ref{fig:grad_norm_overlay} shows the encoder gradient $\ell_2$-norm and
magnitude-spectrogram reconstruction loss recorded during the same 100k-step runs.
The complex STFT paradigm maintains the flattest gradient-norm trajectory with the
least variance and converges to the lowest magnitude loss under this shared training
protocol, complementing its full-band and high-frequency reconstruction results.

\section{Training Data and Quality Pipeline}
\label{app:data}

The full-scale \ours{} is trained on approximately two million publicly sourced
tracks, including LAION-DISCO-12M~\citep{laion2024disco12m,lanzendorfer2023disco10m},
and approximately 10{,}000 hours
of proprietary in-house 48\,kHz stereo music. Detailed source composition is withheld
for data-strategy and licensing reasons. Before training, all sources pass through an
automated quality and provenance pipeline. Raw collections may contain lossless
containers transcoded from lossy sources, duplicated mono tracks labeled as stereo,
clipped masters, and upsampled files with empty high-frequency bands. The pipeline
(Table~\ref{tab:pipeline_rules}) checks encoder tags, metadata, bitrate statistics,
loudness, true peaks, left--right difference energy, and spectral cutoffs.

\subsection{Noise-floor-aware cutoff detection}
Container metadata alone cannot distinguish true high-resolution content from an
upsampled file. We estimate the noise floor from track \emph{edges} and detect the
spectral cutoff from \emph{interior} segments, keeping the two independent. We reserve
the first and last 10\% of the mono waveform as edges and compute Hann-windowed
8192-point RFFTs over non-overlapping windows.

\begin{figure}[h]
  \centering
  \includegraphics[width=\textwidth]{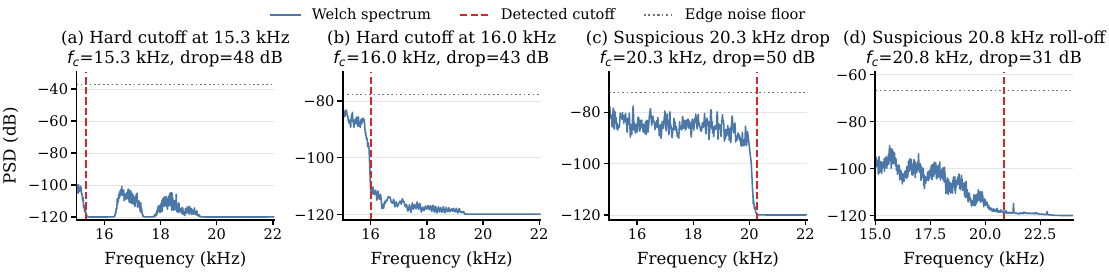}
  \caption{High-frequency cutoff detector examples. Blue: Welch spectrum of a
  verification segment; red dashed: detected cutoff $f_c$; gray dotted:
  independently estimated edge noise floor $\hat\eta$.}
  \label{fig:cutoff}
\end{figure}

Define the peak-normalized magnitude in decibels as
\begin{equation}
  M_{\mathrm{dB}}(f)
  =
  20\log_{10}
  \frac{\bar M(f)}
       {\max_{f'} \bar M(f')+\epsilon_M},
  \qquad
  \epsilon_M>0.
\end{equation}
The empirical noise floor is
\begin{equation}
  \hat\eta
  =
  \operatorname{median}
  \left(
    M_{\mathrm{dB,edge}}[\text{top 10\% bins}]
  \right).
\end{equation}
The detected cutoff $f_c$ is the highest bin whose mean interior magnitude stays
above $-80$\,dB relative to the peak; the local roll-off is
\begin{equation}
  \Delta_{\mathrm{dB}}
  =
  \max M_{\mathrm{dB,int}}
  [f_c{-}2\text{kHz}:f_c]
  -
  \max M_{\mathrm{dB,int}}
  [f_c:f_c{+}2\text{kHz}].
\end{equation}
We replace the post-cutoff maximum with $\hat\eta$ when no reliable content exists
above $f_c$. A file is \texttt{hard\_cutoff} when $f_c/f_\text{Nyq}<0.85$ and
$\Delta_\text{dB}\ge40$\,dB (or $\ge25$\,dB near $\hat\eta$), \texttt{suspicious}
when $f_c/f_\text{Nyq}<0.90$ with $\Delta_\text{dB}\ge20$\,dB, and \texttt{natural}
otherwise (Figure~\ref{fig:cutoff}).

\begin{table*}[tbh]
  \centering
  \scriptsize
  \setlength{\tabcolsep}{3.5pt}
  \renewcommand{\arraystretch}{1.10}
  \caption{Metric-level checklist used by the acoustic pipeline.
  Each row pairs a rule-engine identifier with a brief description
  and its pass/flag condition.}
  \label{tab:pipeline_rules}
  \resizebox{\textwidth}{!}{%
  \begin{tabular}{@{}p{1.8cm}p{3.0cm}p{5.5cm}p{4.3cm}@{}}
    \toprule
    \textbf{Check} & \textbf{Metric ID} & \textbf{Evidence} & \textbf{Condition} \\
    \midrule
    \multirow{2}{1.8cm}{Meta \& encoder}
      & \texttt{encoder\_tag}     & Encoder string in container/stream tags        & flag if \texttt{lavf/lavc/ffmpeg/lame} \\
      & \texttt{mutagen\_tags}    & Metadata sweep for conversion traces           & flag if \texttt{transcode/convert} \\
    \midrule
    \multirow{5}{1.8cm}{Bitrate}
      & \texttt{has\_non\_monotonic\_pts} & PTS monotonicity              & pass if \texttt{false} \\
      & \texttt{bitrate\_utilization}     & Realized / nominal bitrate    & pass if $\geq0.6$ \\
      & \texttt{top10\_bitrate\_mean}     & Complex-segment mean bitrate  & pass if $\geq1.2\times$\,mean \\
      & \texttt{vbr\_variation}           & Bitrate coefficient of variation & pass if $\sigma/\mu\geq0.15$ \\
      & \texttt{quality\_tier}            & Bitrate-based quality level & HiRes/High/Med/Low/Trash \\
    \midrule
    \multirow{4}{1.8cm}{Loudness}
      & \texttt{lufs\_i}            & Integrated loudness (EBU R128)          & $[-25,-5]$\,LUFS \\
      & \texttt{true\_peak}         & dBTP level                              & $[-1,+2]$\,dBFS \\
      & \texttt{lra}                & Loudness range                          & $[3,15]$\,LU \\
      & \texttt{plr}                & Peak-to-loudness ratio                  & $\geq6$\,LU \\
    \midrule
    \multirow{1}{1.8cm}{Fake stereo}
      & \texttt{lr\_diff\_zero\_ratio} & Fraction of zero $|L-R|$ samples & pass if $<0.99$ \\
    \midrule
    \multirow{3}{1.8cm}{Freq cutoff}
      & \texttt{cutoff\_ratio}     & Cutoff / Nyquist freq ratio             & ranking from $0.5 \to 1.0$ \\
      & \texttt{drop\_db}          & Energy gap around $f_c$                 & hard $\geq40$\,dB; suspicious $\geq20$\,dB \\
      & \texttt{valid\_segments}   & Non-silent interior FFT windows         & pass if $>0$ \\
    \bottomrule
  \end{tabular}}
\end{table*}

\section{Metric Definitions}
\label{app:metrics}

This appendix gives the metric details omitted from the main text
(\S\ref{sec:metrics}).
Throughout this appendix, $S$ (with hats for reconstructions) denotes a generic spectrogram in metric formulas, distinct from the native-model STFT $\mathbf{X}$ and the per-channel refiner spectrogram $X$.

\paragraph{\stftdist{} and \logstftdist{}.}
We follow the STFT magnitude distance in \texttt{auraloss}~\citep{steinmetz2020auraloss}
(\texttt{STFTMagnitudeLoss}, L1, mean reduction), not spectral convergence.
Spectral convergence is the separate Frobenius-ratio term
$\lVert |S|-|\hat S|\rVert_F/\lVert |S|\rVert_F$; it appears in the training
multi-resolution STFT loss (\S\ref{sec:losses}) but is not a reported evaluation metric.
Let $|S_k|$ and $|\hat S_k|$ denote reference and reconstructed magnitudes at scale $k$,
with $n_\text{fft}\in\{64,128,256,512,1024,2048\}$, hop $=n_\text{fft}/4$, a Hann window,
and $N_k=F_kT_k$ time--frequency bins. Averaging over $K$ resolutions,
\begin{equation}
  \mbox{\stftdisttab{}}
    =\frac{1}{K}\sum_k\frac{1}{N_k}\big\lVert |S_k|-|\hat S_k|\big\rVert_1.
\end{equation}
\textbf{\logstftdist{}} uses that same L1 magnitude distance after
$\operatorname{log1p}(x)=\log(1+x)$, matching auraloss log compression with
$\texttt{log\_fac}{=}1$ and $\texttt{log\_eps}{=}1$:
\begin{equation}
  \mbox{\logstftdisttab{}}
    =\frac{1}{K}\sum_k\frac{1}{N_k}
      \big\lVert \operatorname{log1p}(|S_k|)-\operatorname{log1p}(|\hat S_k|)\big\rVert_1.
\end{equation}

\paragraph{\meldist{} and \logmeldist{}.}
\textbf{\meldist{}} and \textbf{\logmeldist{}} replace $|S_k|$ with a 64-bin mel spectrogram
and otherwise use the same two auraloss magnitude L1 distances,
abbreviated \meldisttab{} and \logmeldisttab{}.

\paragraph{Spectral Pan Error (SPE).}
\label{app:spe}
The per-bin, per-frame pan position is defined via the intensity
panning law~\citep{avendano2002stereo},
\begin{equation}
  \text{pan}(f,t)=\frac{|L(f,t)|^2}{|L(f,t)|^2+|R(f,t)|^2+\epsilon_\text{pan}},\quad \epsilon_\text{pan}=10^{-8}.
  \label{eq:pan}
\end{equation}
At each scale, left and right channel powers are computed from Hann-windowed complex
STFTs. SPE is the mean L1 deviation between reconstructed
and reference pan profiles over all $N_k=F_kT_k$ time--frequency bins, averaged over
$K{=}4$ STFT scales ($n_\text{fft}\in\{512,1024,2048,4096\}$):
\begin{equation}
  \text{SPE}=\frac{1}{K}\sum_{k}\frac{1}{N_k}
    \sum_{f,t}\big|\,\widehat{\text{pan}}_k(f,t)-\text{pan}_k(f,t)\,\big|,
  \label{eq:spe}
\end{equation}
where $\widehat{\text{pan}}_k$ and $\text{pan}_k$ denote reconstructed and reference
pan profiles, respectively. The additive $\epsilon_\text{pan}$ stabilizes the ratio near zero
energy and is distinct from $\epsilon_\text{num}$ and $\epsilon_\text{init}$ in \specsnake{}; no silence mask is applied before averaging.

\section{Exact-Alignment STFT Wrapper}
\label{app:stft_wrapper}

Let $L$ be the waveform length, $N$ the FFT size, and $H$ the hop size. For waveforms whose length $L$ is divisible by the hop size $H$, preserving the $1920\times$ temporal ratio requires exactly $T=L/H$ STFT frames. PyTorch's default centered analysis produces
$1+\lfloor L/H\rfloor=T+1$ frames. Removing the extra frame to meet the network
interface also removes one hop of synthesis support at the boundary, so the waveform
must be shortened or padded rather than reconstructed at its original length. We
therefore avoid this frame-count--boundary-coverage trade-off.

Disabling implicit padding does not by itself solve the problem. A Hann-windowed
\texttt{center=False} inverse must satisfy the Nonzero-Overlap-Add (NOLA) condition at
every finite-signal boundary: the sum of squared, overlapping synthesis-window values
must be nonzero at every reconstructed sample. Because the Hann window vanishes at its endpoint,
PyTorch's boundary check rejects the unpadded configuration used here
($N{=}960$, $H{=}480$), blocking a differentiable STFT--iSTFT path. Likewise, mixing
center conventions without an explicit boundary construction changes the temporal
origin. We instead define that construction directly in
Algorithm~\ref{alg:stft_wrapper}.

\begin{algorithm}[h]
  \caption{Exact-alignment STFT--iSTFT wrapper.}
  \label{alg:stft_wrapper}
  \begin{algorithmic}[1]
    \Require Waveform $x\in\mathbb{R}^{B\times L}$; FFT size $N$; hop $H$;
      window length $W$; window $w$
    \Ensure Spectrum with $T=L/H$ frames and reconstruction $\hat{x}$ of length $L$
    \Procedure{Analysis}{$x,N,H,W,w$}
      \State $p_\ell\gets N/2$, $p_r\gets H/2$
      \State $x_{\mathrm{pad}}\gets\Call{ZeroPad}{x,\,p_\ell,\,p_r}$
      \State $X\gets\Call{STFT}{x_{\mathrm{pad}},N,H,W,w,\,\texttt{center=False}}$
      \State \Return $X$
    \EndProcedure
    \Procedure{Synthesis}{$X,N,H,W,w,L$}
      \State $p_r\gets H/2$
      \State $\tilde{x}\gets\Call{iSTFT}{X,N,H,W,w,\,\texttt{center=True},
        \texttt{length}=L+p_r}$
      \State \Return $\tilde{x}[:,0{:}L]$
    \EndProcedure
  \end{algorithmic}
\end{algorithm}

\begin{figure}[h]
  \centering
  \includegraphics[width=0.92\textwidth]{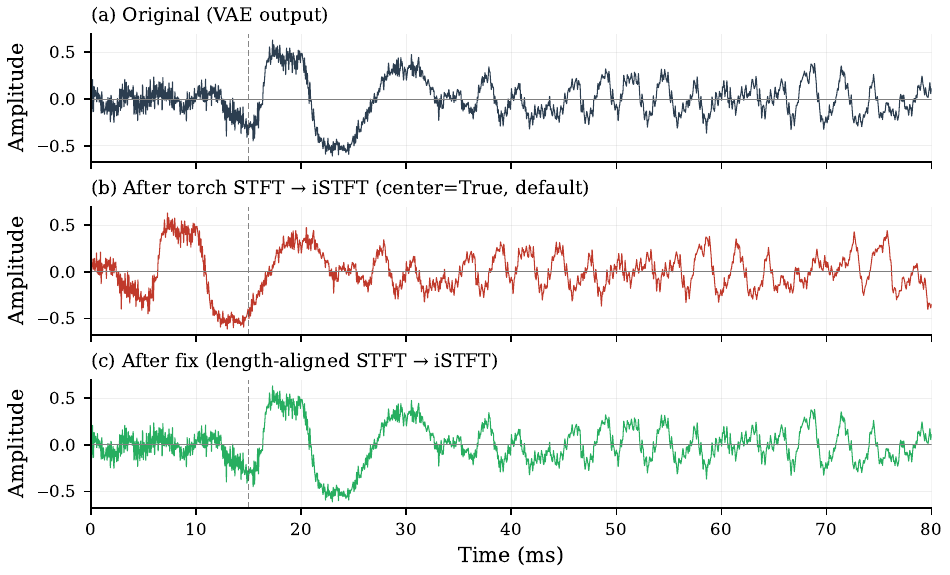}
  \caption{Sample-level comparison over 80\,ms at 48\,kHz; the dashed line marks the
  same reference instant in every panel. \textbf{(a)}~Original waveform.
  \textbf{(b)}~A naive mixed-center round-trip (analysis
  \texttt{center=False}, synthesis \texttt{center=True}) shifts the waveform left by
  one hop (${\approx}10$\,ms). A fully uncentered Hann round-trip is instead rejected
  by the boundary NOLA check and therefore has no reconstructed trace.
  \textbf{(c)}~Our explicitly padded wrapper produces $T$ frames and recovers
  sample-level alignment.}
  \label{fig:stft_drift}
\end{figure}

For our $N{=}2H$ setting, the asymmetric analysis padding gives
$1+\lfloor(L+N/2+H/2-N)/H\rfloor=L/H$ frames. The left pad moves original sample
zero into a nonzero-overlap region and is removed by the centered inverse; requesting
$L+H/2$ synthesis samples supplies the explicit tail margin, which is then cropped.
The wrapper therefore returns exactly $L$ aligned samples, avoids the rejected NOLA
boundary, and remains fully differentiable. Figure~\ref{fig:stft_drift} shows the
sample-level effect.

\section{Convolution-Adapted Muon Optimizer}
\label{app:muon}

The Muon optimizer~\citep{jordan2024muon} orthogonalizes each momentum update with a quintic
Newton--Schulz iteration. It was designed for the 2D weight matrices used in
transformers. To apply it to convolutional spectral networks, we flatten every
${\geq}2$D kernel into an output-channel-by-fan-in matrix, orthogonalize the update, and
reshape it to the original kernel dimensions.

The Muon optimizer updates all ${\geq}2$D convolutional and linear weights except output projections
and discriminator post-convolutions. These output layers remain on the internal AdamW
branch, together with bias and normalization parameters. We retain the standard
$0.2\sqrt{\max(A,B)}$ shape-dependent learning-rate scaling. Stages~1--2 use the Muon optimizer to
reduce gradient spikes during adversarial training and to improve reconstructed
high-frequency detail. Stage~3 uses AdamW for the refiner.

\begin{figure}[h]
  \centering
  \includegraphics[width=0.9\textwidth]{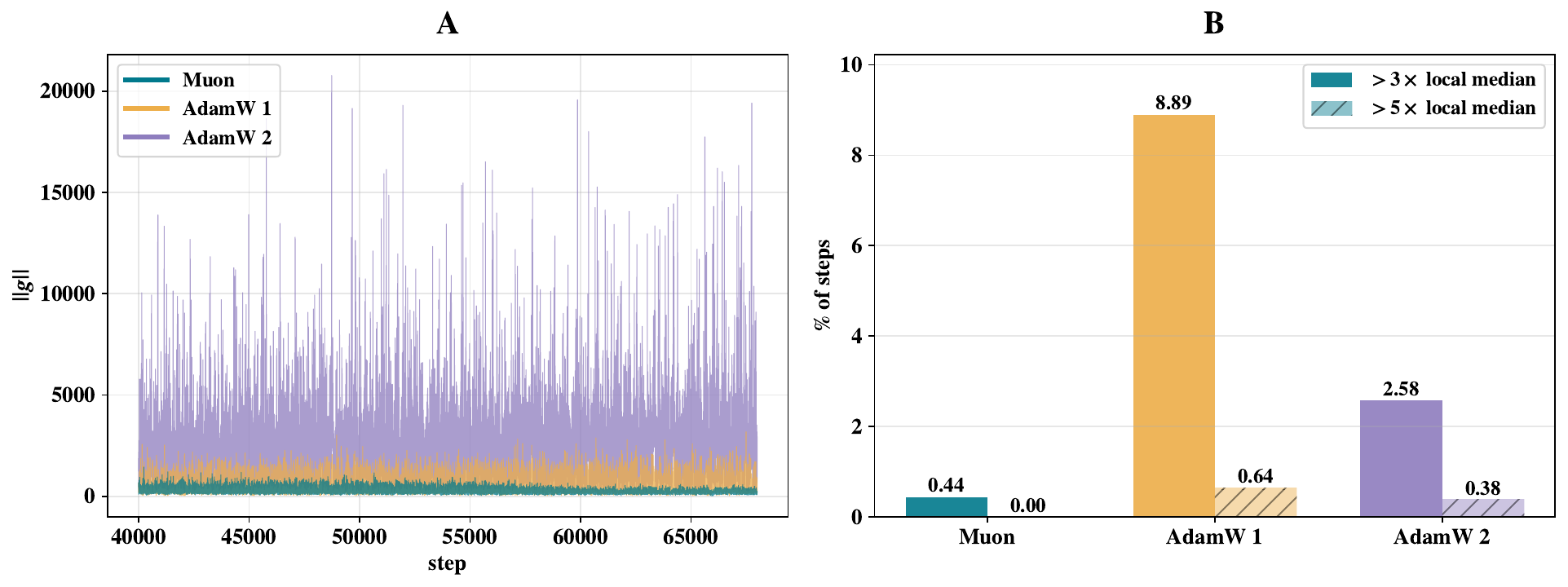}
  \caption{Generator gradient-norm dynamics for Muon and AdamW, measured over a
  common training window (steps 40k--68k) in which all three runs are stable.
  \textbf{(A)}~Per-step generator gradient norm $\|g\|$ for Muon and two AdamW runs
  (AdamW~1, AdamW~2), shown on a shared linear scale; Muon and AdamW~2 use the same
  global batch size and AdamW~1 uses a larger batch. \textbf{(B)}~Local spike rate:
  the percentage of steps at which $\|g\|$ exceeds $3\times$ (solid) or $5\times$
  (hatched) its trailing 500-step median.}
  \label{fig:gradnorm_stability}
\end{figure}

\section{Architecture, Hyperparameters, and Training Configuration}
\label{app:arch}
\label{app:training_config}

The production \ours{} comprises a Spec Encoder with base width $C_0{=}256$
(${\approx}549$M parameters), a Spec Decoder with $C_0{=}128$
(${\approx}134.8$M), and a banded \refiner{} (${\approx}8.2$M), totaling
${\approx}692$M parameters. The STFT front end employs $n_\text{fft}{=}960$ ($\Delta f{=}50$\,Hz at $48$\,kHz),
$\text{hop}{=}480$, and the exact-alignment wrapper of
Appendix~\ref{app:stft_wrapper}. The latent is 128-dimensional at 25\,Hz. The refiner is a
12-layer, 256-dimensional ConvNeXt-1D backbone reading a 960-dimensional per-frame
real--imaginary spectral vector and producing $532$ per-frame residual dimensions per channel (1064
across the stereo pair) with the band assignment described in \S\ref{sec:refiner}.
Training uses 1.28\,s segments; the three-stage schedule and loss weights are
specified in Table~\ref{tab:training_config}. For the generator in Stages~1--2, we
employ the Muon optimizer (momentum orthogonalized by a quintic Newton--Schulz iteration)
with implementation adaptations for the stereo spectral setting.
All \specsnake{} layers use the F-Log configuration: layer $\ell$ learns one
$\alpha_f,\beta_f$ pair for each of its $F_\ell$ frequency positions and shares it across feature channels. We store $\tilde\alpha$ and $\tilde\beta$ and recover $\alpha=e^{\tilde\alpha}$, $\beta=e^{\tilde\beta}$.

Table~\ref{tab:training_config} summarizes the three-stage schedule.
The multi-resolution STFT loss uses seven FFT sizes from 32 to 2048. The adversarial
stack contains an eight-scale STFT discriminator with per-channel
$[\operatorname{Re}(X),\operatorname{Im}(X),|X|]$ inputs. Five scales use
$n_\text{fft}\in\{128,256,512,1024,2048\}$ with proportionally scaled hops; three
scales use $n_\text{fft}{=}4096$ with hops 2048, 1024, and 256. The stack also includes
three CQT sub-discriminators at 24, 36, and 48 bins per octave over nine octaves.
Stage~3 additionally applies four-scale adaptive frequency pooling to the
refiner's native complex STFT output. Training runs on one node with 8 NVIDIA A100 GPUs; optimizer state is reset between stages.

\begin{table}[h]
    \centering
    \caption{\ours{} training configuration across three stages. Stage~1 is
    reconstruction-only; Stage~2 adds waveform-domain adversarial training;
    Stage~3 freezes the encoder--decoder and trains the refiner with $D_w$ and $D_s$.
    Adv/FM denotes adversarial (Adv) / feature-matching (FM) loss weights.}
    \label{tab:training_config}
    \begin{tabular}{@{}llccc@{}}
        \toprule
        Category & Param & Stage~1 & Stage~2 & Stage~3 \\
        \midrule
        \multirow{4}{*}{Optimizer}
          & Generator optim.  & Muon      & Muon                & AdamW \\
          & Generator LR      & $10^{-4}$ & $1.5{\times}10^{-4}$ & $1.5{\times}10^{-4}$ \\
          & Disc.\ optim.     & ---       & Muon                & AdamW \\
          & Disc.\ LR         & ---       & $3{\times}10^{-5}$   & $10^{-6}$($D_w$)\,/\,$10^{-5}$($D_s$) \\
        \midrule
        \multirow{5}{*}{Schedule}
          & Training duration & 1 epoch   & 5 epochs   & 3 epochs \\
          & Batch size / GPU  & 4       & 1        & 1 \\
          & Segment length    & 1.28\,s & 1.28\,s  & 1.28\,s \\
          & Disc.\ warmup     & ---     & 0        & 0($D_w$)\,/\,15k($D_s$) \\
          & G:D interleave    & ---     & 1:1      & 4:1 \\
        \midrule
        \multirow{3}{*}{Disc.}
          & STFT disc.\ scales & --- & 8          & 8 \\
          & CQT disc.          & \texttimes & \checkmark & \checkmark \\
          & Spectral disc.     & \texttimes & \texttimes & \checkmark \\
        \midrule
        \multirow{5}{*}{\makecell[l]{Loss\\weights}}
          & $\lambda_\text{STFT}$              & 1.0       & 1.0       & 0.5 \\
          & $\lambda_\text{IF/GD}$             & 0         & 0         & 0.1 \\
          & $\lambda_\text{KL}$                & $10^{-6}$ & $10^{-6}$ & 0 \\
          & $\lambda_\text{time-GAN}$ (Adv/FM) & ---       & 1.0/100   & 0.175/20 \\
          & $\lambda_\text{spec-GAN}$ (Adv/FM) & ---       & ---       & 0.5/35 \\
        \bottomrule
    \end{tabular}
\end{table}

\section{Why a Spectral-Domain Discriminator in Stage~3?}
\label{app:spec_disc}

This section is a \emph{diagnostic exploration} of the Stage~3 design, not a core
claim. All probes are training-free (forward and backward passes only; no weights are
updated) and use our trained generator and discriminators; probe inputs are 4\,s
excerpts from our internal evaluation set. The structural quantities of
\S\ref{app:spec_disc_blind} depend only on the STFT configuration, not on model
capacity.

\subsection{Starting point: supervision at the Stage~3 boundary}
\label{app:spec_disc_start}

The first two rows of Table~\ref{tab:spec_disc_sep} probe $D_w$ at the ends of
Stages~2 and~3, respectively. At the Stage~2 checkpoint, its mean real and fake logits are $0.62$ and $0.41$,
respectively, leaving a gap of $0.21$. This gap does not establish equilibrium, but
$D_w$ has already been trained on these outputs. Stage~3 then attaches the refiner,
whose zero-initialized residuals make its initial output equal to the Stage~2 decoder
output (\S\ref{sec:refiner}). We therefore test whether direct spectral supervision
provides output-space gradients that are unavailable through the iSTFT path.

\begin{table}[t]
  \centering
  \small
  \caption{Discriminability probe at the Stage~3 boundary ($n{=}48$ segments): mean
  real/fake logits and real--fake gap. $D_w^{(2)}$/$D_w^{(3)}$ are $D_w$
  at the end of Stage~2 and after Stage~3 continuation; $D_s$ is the trained
  spectral-domain discriminator; $D_s{\circ}P$ scores inputs pre-projected onto the
  consistent subspace (gauge component removed); $D_s^{\text{rand}}$ is a randomly
  initialized spectral-domain discriminator.}
  \label{tab:spec_disc_sep}
  \begin{tabular}{@{}lcccl@{}}
    \toprule
    Critic & logit(real) & logit(fake) & gap & Note \\
    \midrule
    $D_w^{(2)}$         & 0.619 & 0.409 & 0.210 & Stage~2 checkpoint \\
    $D_w^{(3)}$         & 0.551 & 0.366 & 0.186 & after Stage~3 continuation \\
    $D_s$               & 0.485 & 0.034 & \textbf{0.451} & trained spectral \\
    $D_s{\circ}P$      & 0.485 & 0.449 & 0.036 & gauge removed \\
    $D_s^{\text{rand}}$ & $-$0.003 & $-$0.002 & 0.001 & random init \\
    \bottomrule
  \end{tabular}
\end{table}

\subsection{Making the conjecture precise: the synthesis-null subspace}
\label{app:spec_disc_blind}

With $n_\text{fft}{=}960$ and hop $480$, the one-sided complex spectrogram has approximately twice as many real output coordinates as the waveform representation under this STFT configuration. The exact finite-signal ratio
depends on boundary handling and on the real constraints at the DC and Nyquist bins,
but the synthesis-null subspace occupies approximately half of the real output
dimension. Define the round-trip operator
$P=\mathrm{STFT}\circ\mathrm{iSTFT}$, which is a near-orthogonal projector onto the
consistent-spectrogram subspace. Any differentiable objective that
reaches a spectrogram $S$ only through $w=\mathrm{iSTFT}(S)$ has zero directional
derivative along $\ker(\mathrm{iSTFT})$, regardless of the transform applied by the
critic. Perturbations in this kernel leave the resynthesized waveform unchanged. They
are redundant directions of the representation, not lost acoustic information. Thus,
waveform-routed supervision cannot constrain these directions.

\subsection{What the spectral discriminator supervises}
\label{app:spec_disc_probe}

Three probes characterize what $D_s$ responds to at the Stage~3 boundary.

\paragraph{The null subspace is populated.}
The Stage~2 output lies measurably outside the consistent subspace: its
inconsistency $\|S-PS\|/\|S\|$ is $0.234\pm0.035$ ($n{=}48$), i.e.,\ about $5.5\%$ of the squared spectral norm lies in the null component that waveform-routed objectives cannot
differentiate.

\paragraph{The discriminator's gradient concentrates in the null subspace.}
Table~\ref{tab:spec_disc_grad} dissects the generator-side gradients (adversarial
plus feature-matching terms, weighted as in training) with respect to $S$. Gradients
routed through the iSTFT retain only $0.02\%$ of their energy in the null subspace,
which is the predicted zero up to boundary effects. In contrast, $95.3\%$ of the
spectral discriminator's gradient energy lies inside it. The two paths are nearly
orthogonal: $\cos(g_w^{(3)},g_s)=-0.002\pm0.003$ over the full space and
$-0.012\pm0.013$ within the consistent subspace ($n{=}24$).

\begin{table}[t]
  \centering
  \small
  \caption{Gradient anatomy at the Stage~3 boundary ($n{=}24$): fraction of each
  generator-side gradient's energy inside the synthesis-null subspace.
  $g_w^{(2)},g_w^{(3)}$ are waveform-path gradients (through the iSTFT) from
  Stage~2 and Stage~3 $D_w$; $g_s$ is $D_s$'s direct
  gradient.}
  \label{tab:spec_disc_grad}
  \begin{tabular}{@{}lc@{}}
    \toprule
    Gradient & null-space energy fraction \\
    \midrule
    $g_w^{(2)}$ (Stage~2 $D_w$) & 0.0002 \\
    $g_w^{(3)}$ (Stage~3 $D_w$) & 0.0002 \\
    $g_s$ ($D_s$, direct)       & \textbf{0.9529} \\
    \bottomrule
  \end{tabular}
\end{table}

\paragraph{Selectivity controls.}
We tested the alternative reading in which $D_s$ mainly provides a finer content
view. The controls disfavor this as the dominant explanation. Pre-projecting its
inputs onto the consistent subspace
collapses the trained discriminator's real--fake gap from $0.45$ to $0.04$
(Table~\ref{tab:spec_disc_sep}), while a randomly initialized spectral discriminator
shows no separation at all (gap $0.001$): the margin is learned, and it is learned
from the gauge component. Sensitivity probes point the same way. We define $z$ as the
absolute mean logit change divided by the baseline logit standard
deviation. Pure null-space perturbations that leave the waveform unchanged (relative
change $\le2.4{\times}10^{-5}$) are invisible to $D_w$ ($z{=}0.00$) yet move
$D_s$ by $z{=}5.1$ at a $1\%$ spectral norm ratio and $z{=}87$ at $30\%$. Four
waveform-calibrated content-artifact families (additive noise, phase jitter, spectral
ripple, phase walk; relative waveform change $0.1\%$--$10\%$) produce $z\le0.16$
for both $D_w$ and $D_s$. A per-position mean $|\Delta\text{logit}|$ analysis gives
the same ordering, excluding sign cancellation.

\subsection{Trajectory during Stage~3 training}
\label{app:spec_disc_traj}

\begin{figure}[t]
  \centering
  \includegraphics[width=0.7\textwidth]{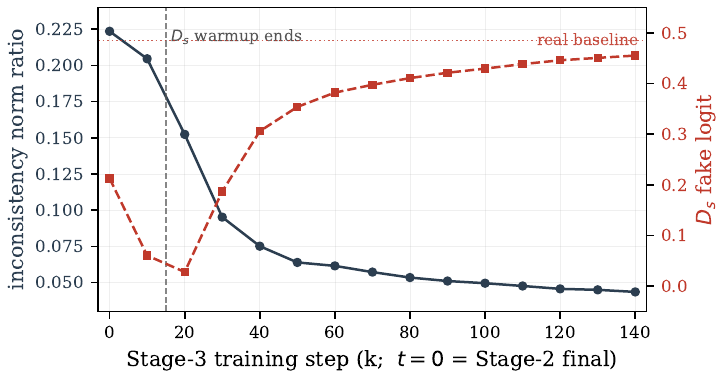}
  \caption{Stage~3 consistency trajectory ($n{=}18$ segments), scored by the fixed
  final spectral discriminator $D_s$. \textbf{Left axis}: inconsistency
  $\|S{-}PS\|/\|S\|$ of the refiner output ($t{=}0$ is the Stage~2 final generator).
  \textbf{Right axis}: $D_s$'s fake logit; the dotted line marks its real baseline
  ($0.485$). The steepest descent follows the end of $D_s$'s warmup. The
  coincident spectral-metric improvement is correlational, not an attribution
  (\S\ref{app:spec_disc_traj}).}
  \label{fig:spec_disc_traj}
\end{figure}

Figure~\ref{fig:spec_disc_traj} tracks the inconsistency of the refiner output across
Stage~3 checkpoints, scored by the fixed final spectral discriminator. At the measured
checkpoints, the null-space component decreases from $0.224$ to $0.044$, a fivefold
reduction. The largest declines occur in the first checkpoint intervals after the
discriminator's warmup ends. The discriminator's fake logit on the refiner output, $0.03$
at the boundary (Table~\ref{tab:spec_disc_sep}), rises to $0.46$ by the end of
training, against a real baseline of $0.485$: the gauge-driven separability that
marked the Stage~3 boundary disappears as training proceeds. Over the same window the
spectral evaluation metrics (\meldist{} and \stftdist{}) improve sharply. We report
this as a temporal coincidence, not an attribution. The direct spectral reconstruction
loss is active throughout Stage~3 and also shapes the null directions. Isolating the
discriminator's causal contribution therefore requires a controlled training arm.

\subsection{Summary and limitations}
\label{app:spec_disc_summary}

At the Stage~3 boundary, $D_w$ retains a real/fake logit gap of $0.21$ on the outputs
that the identity-initialized refiner reproduces. The redundant spectral
parameterization has an approximately half-dimensional subspace that waveform-routed
objectives cannot observe, and the Stage~2 outputs contain energy in this subspace. In
our probes, $95.3\%$ of the spectral discriminator's output-gradient energy lies in
these directions, nearly orthogonal to the waveform path. The measured null component
then falls fivefold during training. These results suggest that $D_s$ acts mainly as a
learned regularizer of waveform-unidentifiable spectral inconsistency (gauge fixing),
rather than as a source of complementary audible-content discrimination.

Several limitations apply. The probes use a small evaluation set with $18$--$48$
segments per condition. Sensitivity is summarized only through mean logits and
per-position mean $|\Delta\text{logit}|$. The evidence is diagnostic: it identifies
the signal that $D_s$ uses in these probes but does not establish the size of its
causal contribution. That question requires the controlled training arms discussed in
\S\ref{app:spec_disc_traj}.

\section{\specsnake{}: Formulation and Initialization Analysis}
\label{app:specsnake}

This appendix separates exact algebraic properties from architectural design choices.
Snake adds a periodic residual in the feature-value coordinate $x$; SnakeBeta
separately parameterizes the residual period and range; and \specsnake{} indexes these
parameters by spectral position while sharing them across feature channels. The first two steps are algebraic,
whereas the frequency indexing and its initialization are architectural priors whose
utility is evaluated empirically. Throughout, $x$ denotes a scalar pre-activation. We
optimize $\tilde\alpha$ and $\tilde\beta$ in log space, with $\alpha=e^{\tilde\alpha}$ and $\beta=e^{\tilde\beta}$, and use
$\epsilon_\text{num}{=}10^{-9}$ in the denominator and
$\epsilon_\text{init}{=}10^{-6}$ in the initialization.

A finite ReLU network on $\R$ defines a piecewise-linear (PWL) map with finitely many
pieces. Any such map that is globally periodic must be constant; hence a finite ReLU
network cannot exactly represent a nonconstant periodic function on the entire real
line~\citep{ziyin2020snake}. On compact intervals, however, ReLU networks can
approximate periodic functions, so this observation motivates an explicit periodic
basis rather than establishing its necessity. Importantly, it concerns periodicity in
the feature-value coordinate $x$, not harmonic spacing along the spectral-frequency
coordinate $f$. We therefore use Snake's periodic residual as an audio inductive bias,
and use the explicit STFT frequency axis for a separate purpose: allowing this bias to
vary with spectral position.

Snake adds a periodic residual to an identity skip,
\begin{equation}
  \text{snake}_\alpha(x)=x+\tfrac1\alpha\sin^2(\alpha x)
    =x+\tfrac{1-\cos 2\alpha x}{2\alpha},
\end{equation}
where $\alpha>0$. Its residual has period $\pi/\alpha$ and range
$[0,1/\alpha]$, while the complete activation has derivative
$1+\sin(2\alpha x)\in[0,2]$ and is therefore monotone. Moreover, for fixed $x$,
\begin{equation}
  \tfrac1\alpha\sin^2(\alpha x)
    = \alpha x^2 + O(\alpha^3x^4),
    \qquad \alpha\to0,
\end{equation}
so the activation converges pointwise to the identity. As the period $\pi/\alpha$ and
the range $1/\alpha$ show, Snake couples the residual period and its maximum range
through the same parameter $\alpha$.

SnakeBeta introduces $\beta>0$ so that the residual period $\pi/\alpha$ and maximum
range $1/(\beta+\epsilon_\text{num})$ are separately parameterized. This separation
removes Snake's unconditional monotonicity guarantee. Differentiating the implemented
activation gives
\begin{equation}
  \tfrac{d}{dx}\!\left[x+\tfrac1{\beta+\epsilon_\text{num}}\sin^2(\alpha x)\right]
    = 1+\tfrac{\alpha}{\beta+\epsilon_\text{num}}\sin(2\alpha x)
    \ \in\ \Big[1-\tfrac\alpha{\beta+\epsilon_\text{num}},\
                   1+\tfrac\alpha{\beta+\epsilon_\text{num}}\Big].
\end{equation}
The derivative is non-negative for all $x$ if and only if
$\alpha\le\beta+\epsilon_\text{num}$. This condition is tight because the lower bound
is reached when $\sin(2\alpha x)=-1$. We permit non-monotonicity because hidden
features need not preserve scalar order. The identity skip centers the derivative
interval at one; it does not impose a parameter-uniform derivative bound.

For clarity, we suppress the layer index below and write $F=F_\ell$ and
$\mathrm{freq}_f=\mathrm{freq}_{\ell,f}$ for a given layer. Conventional channel-wise SnakeBeta applies one curve $(\alpha_c,\beta_c)$ to all $F$
spectral positions. \specsnake{} instead learns
$\tilde\alpha,\tilde\beta\in\R^{F}$ (Eq.~\ref{eq:snakebeta}), yielding one curve per
physical frequency and applying it to every feature channel.
We use $C$ for channel-indexed parameters, $F$ for frequency-indexed parameters shared
across channels, and $CF$ for independent channel--frequency parameters. F-Log,
F-Uniform, and CF-Log denote the corresponding initialized configurations. $C$
contains $2C$ activation parameters per layer and is not the left/right audio channel.
$F$ contains $2F$ parameters per layer. $CF$ learns a separate curve for every $(c,f)$ pair
and therefore contains $2CF$ parameters per layer. CF-Log begins from the same
frequency-proportional values as F-Log, replicated across channels, but optimizes the
copies independently. The default \specsnake{} is the channel-shared F-Log
configuration. Its log-space parameterization guarantees positivity and gives
\begin{equation}
  \frac{\partial\mathcal L}{\partial\tilde\alpha_f}
    =\alpha_f\frac{\partial\mathcal L}{\partial\alpha_f},
  \qquad
  \frac{\partial\mathcal L}{\partial\tilde\beta_f}
    =\beta_f\frac{\partial\mathcal L}{\partial\beta_f}.
\end{equation}
These identities describe the reparameterization and do not imply a convergence
advantage.

We initialize $\alpha_f$ in proportion to normalized physical frequency using
Eq.~\ref{eq:alphainit}, so that
\begin{equation}
  \alpha^{(0)}_f
    = \frac{\text{freq}_f}{\bar f}+\epsilon_\text{init}.
\end{equation}
For equally spaced retained bins
$\mathrm{freq}_f=f\Delta$, $f=0,\ldots,F-1$,
the mean frequency is $\bar f=(F-1)\Delta/2$, yielding
\begin{equation}
  \alpha_f^{(0)}
  =
  \frac{2f}{F-1}
  +\epsilon_\mathrm{init}.
\end{equation} 
The effective initialization is thus
linear in normalized frequency, although the positive parameter is stored in log-space.

At initialization, $\tilde\beta_f=0$ and hence $\beta_f=1$. To isolate the
effect of the $\alpha_f$ schedule, consider an analysis slice in which the marginal
pre-activation distribution is $x_{c,f}\sim\mathcal{N}(0,\sigma^2)$ with the same
$\sigma>0$ across frequency bins. Define the residual correction
$\Delta_f(x)=\sin^2(\alpha_f x)/(1+\epsilon_\text{num})$. Using
$\mathbb{E}[\cos(tx)]=e^{-t^2\sigma^2/2}$ with $t=2\alpha_f$ gives
\begin{equation}
  \mathbb{E}[\Delta_f(x)]
    =\frac{1-e^{-2\alpha_f^2\sigma^2}}
           {2(1+\epsilon_\text{num})},
  \label{eq:esin2}
\end{equation}
which is strictly increasing in $\alpha_f$. Under the stated common-distribution
assumption, the expected correction therefore increases with spectral position. This
is a conditional distributional statement, not a pointwise one: if feature means or
variances differ across bins, the ordering need not hold.

At DC, $\alpha^{(0)}_0=\epsilon_\text{init}$ and, for fixed $x$,
\begin{equation}
  \Delta_0(x)
    =\frac{\epsilon_\text{init}^2x^2}{1+\epsilon_\text{num}}
      +O(\epsilon_\text{init}^4x^4),
\end{equation}
so the activation is effectively identity-like at initialization for bounded
pre-activations. In contrast,
$\sup_x\Delta_f(x)=1/(1+\epsilon_\text{num})$ is identical for every $f$.
Thus, the uniform $\beta$ initialization introduces no frequency-dependent maximum
residual range: $\alpha_f$ controls the feature-space oscillation rate, while the
frequency ordering of the expected correction in Eq.~\ref{eq:esin2} depends on the
stated input-distribution assumption.

In summary, the $F$ parameterization assigns a scalar nonlinearity to each physical
frequency and shares it across feature channels; F-Log applies frequency-proportional
initialization in $\alpha$-space while storing and optimizing the positive parameters
in log space. Its initialization orders
feature-space oscillation rates by normalized spectral position while remaining
identity-like near DC\@. In the controlled ablation (\S\ref{sec:ablation_activation}),
F-Log outperforms channel-wise $C$ on five of six metrics and fully independent $CF$
on all six metrics.

\section{Activation Ablation: Protocol}
\label{app:activation}

The activation ablation (\S\ref{sec:ablation_activation}) uses a shared mid-scale
encoder--decoder with base width $C_0{=}32$, bottleneck width $D{=}64$, and approximately 42.6M generator
parameters. Each model trains for 100k steps with the multi-resolution STFT loss,
LSGAN loss, feature-matching (FM), and KL regularization. Training uses 48\,kHz stereo
segments of 1.28\,s with a batch size of 4. The data are drawn from the
LAION-DISCO-12M subset~\citep{laion2024disco12m,lanzendorfer2023disco10m} after the
filtering described in Appendix~\ref{app:data}. We train each activation independently
with seeds 42, 233, and 666. Table~\ref{tab:activation} reports the mean across the
three runs without a confidence interval. All configurations use the same data,
architecture, optimization, and evaluation set. F-Uniform and F-Log share one
parameter pair across all feature channels at each frequency bin; they differ only in
initialization (F-Uniform is zero-initialized; Uniform does not refer to frequency sampling). CF-Log instead learns an independent parameter pair per channel--frequency
combination while retaining the F-Log initialization.
Across the full ablation network, F-Log uses 2,730 activation parameters and CF-Log
uses 350,720. Evaluation uses the same 100 Song Describer Dataset files for every seed and
configuration. For CF-Log, per-channel coherence is 0.892 and CCPC
is 0.885; Table~\ref{tab:activation} reports the latter.

\section{Banded-Refiner Efficiency and Cue-Aligned Evaluation}
\label{app:bandmode}

The Banded and \unconstrainedrefiner{} variants are trained only at the refiner stage from the same frozen decoder
under the same optimization protocol (not the production three-epoch refiner). They
differ in output residual dimensions, so the comparison evaluates the efficiency of the
complete structured design rather than isolating band placement at matched capacity.
As reported in the main text (\S\ref{sec:ablation_bandmode}), the banded design uses
fewer residual dimensions while improving magnitude distances and the duplex-aligned
probes most directly targeted by its constraints. Relative
to the \unconstrainedrefiner{}, it reduces Low IPD, Low ILD, and High ILD errors by 2.4\%,
8.1\%, and 14.7\%, respectively, while High IPD remains within 0.7\%. We complement
these measurements with a controlled paired listening test. Ten professional mixing
and mastering engineers scored both versions of each excerpt on a 1--10 scale. Scores
are mapped to $[0,1]$ as $(s-1)/9$, and the default placement receives the higher mean
paired rating (0.75 vs.\ 0.66).

\subsection{Band-Restricted Duplex-Cue Metrics}
\label{app:duplex_cues}

The duplex-cue columns of Table~\ref{tab:bandmode} measure how faithfully a
reconstruction preserves the two interaural cues that the band allocation targets. Both
metrics are computed from the complex STFTs of the reference and the reconstruction,
truncated to a common signal length and frame count, and both are restricted to a bin
range $[f_0,f_1)$: the Low band runs from DC to the bin of the $1.5$\,kHz refiner edge,
and the High band runs from the bin of the $4$\,kHz edge through the Nyquist bin.
Following Eq.~\ref{eq:pan}, $L(f,t)$ and $R(f,t)$ denote the left and right complex STFT
coefficients over $T$ frames, with hats marking the reconstruction. Neither metric
applies a silence mask.

\paragraph{Interaural phase difference (IPD) error.}
The interaural phase difference at each time--frequency bin is
$\Delta\phi(f,t)=\phi_L(f,t)-\phi_R(f,t)$, where $\phi_L=\arg L$ and $\phi_R=\arg R$.
Phase differences are circular, so the deviation from the reference is reduced modulo
$2\pi$ by taking the principal argument of the corresponding unit complex number, which
maps it into $(-\pi,\pi]$:
\begin{equation}
  \text{IPD}_{[f_0,f_1)}
    =\frac{1}{(f_1-f_0)\,T}\sum_{f=f_0}^{f_1-1}\sum_{t}
      \Big|\arg\exp\!\big(j\,[\,\widehat{\Delta\phi}(f,t)-\Delta\phi(f,t)\,]\big)\Big|.
  \label{eq:ipd}
\end{equation}
Values are in radians and are therefore bounded above by $\pi$. Every time--frequency
bin in the band contributes equally, independent of its energy.

\paragraph{Interaural level difference (ILD) error.}
The ILD is a per-frame quantity: band energies are aggregated across the bins of the
band before the level ratio is taken in decibels,
\begin{equation}
  \text{ILD}_{[f_0,f_1)}(t)=10\log_{10}
    \frac{\sum_{f=f_0}^{f_1-1}|L(f,t)|^2+\epsilon_\text{ild}}
         {\sum_{f=f_0}^{f_1-1}|R(f,t)|^2+\epsilon_\text{ild}},
  \qquad \epsilon_\text{ild}=10^{-8},
  \label{eq:ild}
\end{equation}
and the reported error is the mean absolute deviation from the reference across frames,
\begin{equation}
  \text{ILD-Err}_{[f_0,f_1)}
    =\frac{1}{T}\sum_{t}\big|\,\widehat{\text{ILD}}_{[f_0,f_1)}(t)
      -\text{ILD}_{[f_0,f_1)}(t)\,\big|.
  \label{eq:ilderr}
\end{equation}
Values are in decibels. The additive $\epsilon_\text{ild}$ stabilizes the ratio in
near-silent frames and is distinct from $\epsilon_\text{pan}$ in Eq.~\ref{eq:pan}. Unlike
Eq.~\ref{eq:ipd}, the level cue is energy-aggregated within each band and frame, but
frames are still averaged with equal weight. For both metrics, per-track values are
averaged over the evaluation tracks.

\subsection{Subjective Test Protocol}
\label{app:subjective_protocol}

\paragraph{Participants and equipment.}
Ten professional mixing and mastering engineers (all with ${\geq}5$ years of commercial
credits) participated. Each listener used calibrated studio headphones in a treated
listening environment with ambient noise ${\leq}30$\,dB\,SPL\@. Playback was routed
through a dedicated digital-to-analog (D/A) converter at 48\,kHz/24-bit with the monitoring level calibrated
to 70\,dB\,SPL (K-weighted).

\paragraph{Stimuli and randomization.}
From the Song Describer Dataset evaluation set we sampled 20 excerpts (10\,s each),
stratified across genre (pop, rock, electronic, jazz, classical) and dynamic
characteristics. Each engineer evaluated all excerpts for both the duplex-placement and
\unconstrainedrefiner{} conditions as matched A/B pairs. For each excerpt,
the assignment of the two conditions to A and B and the order of presentation were
randomized independently for each engineer. Condition labels remained hidden throughout
the blind evaluation. A 2-second silent gap separated the two members of each pair.

\paragraph{Rating scale.}
Engineers scored both members of each pair on a 1--10 integer scale according to the rubric in
Table~\ref{tab:subjective_rubric}. The scale assesses four perceptual dimensions:
spectral balance (frequency), dynamics, spatial imaging, and harmonic/timbral quality.
Scores are mapped to $[0,1]$ as $(s-1)/9$ for reporting.

\begin{table}[h]
  \centering
  \small
  \caption{Subjective quality rubric. Engineers assign a single holistic score on a
  1--10 scale; anchor descriptions guide consistent interpretation.}
  \label{tab:subjective_rubric}
  \begin{tabular}{@{}cp{11.5cm}@{}}
    \toprule
    \textbf{Score} & \textbf{Anchor Description} \\
    \midrule
    8--10 & Frequency balance suits the genre; mid/side energy ratio appropriate; no
            masking between instruments; dynamics well controlled with natural
            transients; spatial staging coherent with adequate depth; harmonic
            coloration (if any) enhances rather than muddies. \\
    5--7  & Minor spectral gaps or slight imbalance in one band; compression
            artifacts occasionally audible (pumping, over-limited transients);
            spatial image somewhat flat but without obvious panning errors; timbral
            quality acceptable with minor coloration. \\
    2--4  & Noticeable frequency-band dropout or harshness (e.g.,\ sibilance, muddy
            low-mids); dynamics feel unnatural (over-compressed or under-controlled);
            spatial image collapsed or inconsistent; timbral distortion clearly
            audible. \\
    1     & Severe artifacts across multiple dimensions; reconstruction
            quality unacceptable for professional evaluation. \\
    \bottomrule
  \end{tabular}
\end{table}

\paragraph{Score aggregation.}
Each excerpt was rated by all ten engineers under both conditions. The reported
score per condition is the arithmetic mean of the normalized paired ratings over all
engineers and excerpts.

\section{A Latent Temporal-Frequency Probe (Exploratory)}
\label{app:latent_probe}

As an exploratory investigation of what an audio VAE latent encodes, we probe how
each model distributes information along the \emph{temporal} axis of its latent
sequence. Audio content varies at different rates. Percussion, onsets, and rhythmic
``groove'' change rapidly, whereas sustained harmony and timbre change slowly. A
latent may therefore encode fast and slow temporal changes in different components.
This section constitutes a \emph{diagnostic exploration}, not a core claim.

\paragraph{Probe.}
For each model, we encode a clip to its latent $\mathbf{z}\in\R^{128\times T}$ and take
the real FFT along time. We then split the temporal-frequency bins at the midpoint.
Zeroing the upper half gives a slowly varying latent $\mathbf{z}_\text{low}$; zeroing
the lower half gives a rapidly varying latent $\mathbf{z}_\text{high}$. An inverse FFT
returns each partial latent to the time domain. We decode
$\mathbf{z}_\text{low}$ and $\mathbf{z}_\text{high}$ \emph{separately} and measure the
spectral centroid of each decoded signal. We summarize with an
\emph{inversion ratio} $\rho=\text{centroid}(\mathbf{z}_\text{low}\!\to\!\text{audio})/
\text{centroid}(\mathbf{z}_\text{high}\!\to\!\text{audio})$; $\rho>1$ (``inverted'') means the
rapidly varying latent decodes to lower-centroid audio and the slowly varying latent
decodes to higher-centroid audio.

\begin{table}[h]
  \centering
  \small
  \caption{Latent temporal-frequency probe (spectral centroid of the separately
  decoded low-/high-temporal-frequency latent halves; averaged over our internal evaluation set).
  $\rho>1$ (``inverted'') means the rapidly varying latent decodes to audio with a
  lower spectral centroid. \ours{} and \earvae{} are inverted; the other three models
  are not.}
  \label{tab:latent_probe}
  \begin{tabular}{@{}lcccc@{}}
    \toprule
    Model & centroid$_\text{low}$ (Hz) & centroid$_\text{high}$ (Hz) & $\rho$ & inverted \\
    \midrule
    \ours{} (complex STFT)                 & 323 & 137 & \textbf{2.36} & \checkmark \\
    \earvae{} (waveform)                   & 477 & 249 & 1.82 & \checkmark \\
    LeVo~2                                 & 392 & 736 & 0.53 & \texttimes \\
    SAME-L                                 & 391 & 487 & 0.79 & \texttimes \\
    SA-Open                                & 363 & 393 & 0.91 & \texttimes \\
    \bottomrule
  \end{tabular}
\end{table}

\begin{figure}[h]
  \centering
  \includegraphics[width=\textwidth]{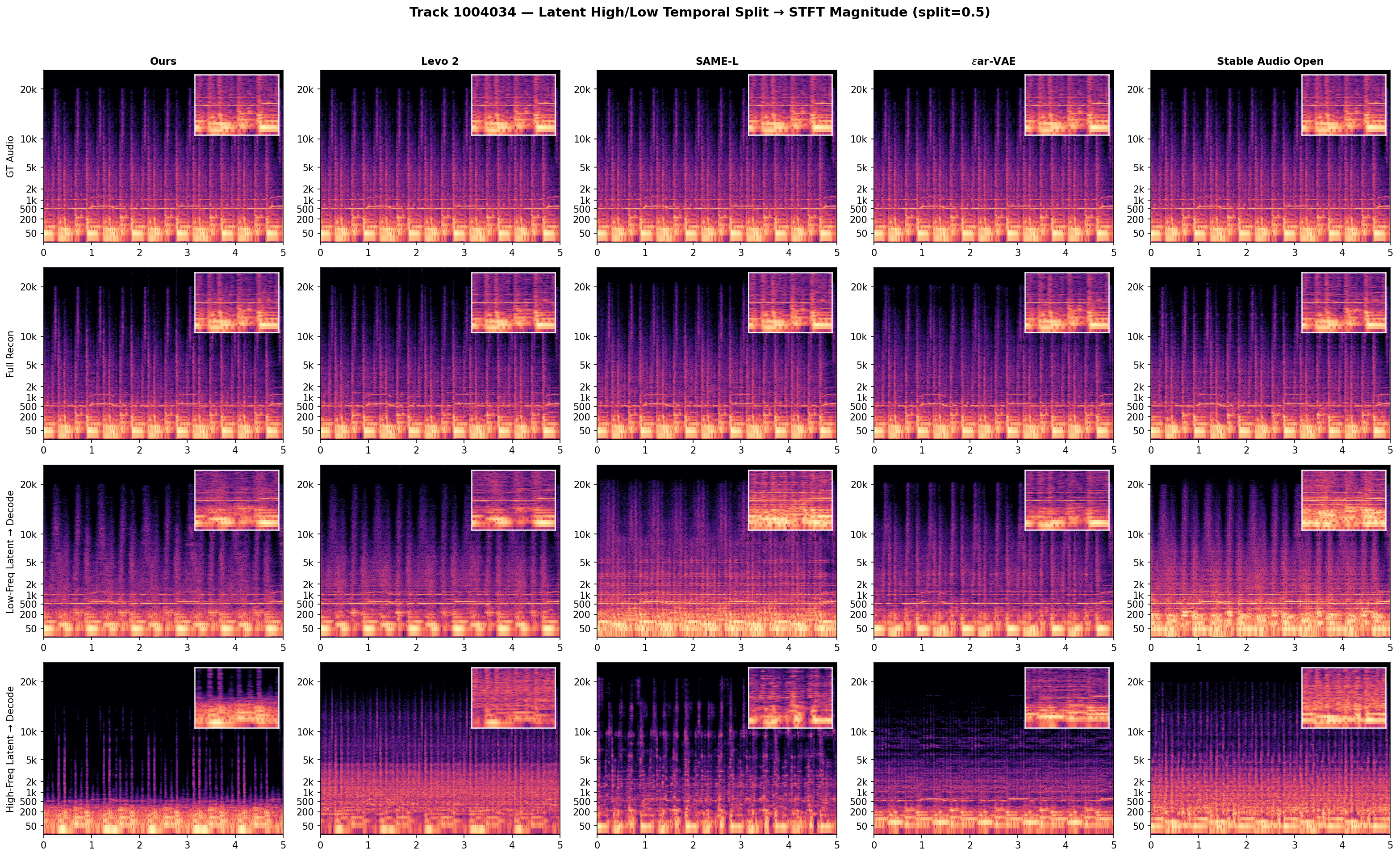}
  \caption{Latent temporal-frequency probe for one track (Track 1004034; temporal
  split at 0.5). \textbf{Rows} (top to bottom): ground-truth audio; full reconstruction;
  low-temporal-frequency latent half decoded; high-temporal-frequency latent half
  decoded. \textbf{Columns} (left to right): \ours{}, LeVo~2, SAME-L, \earvae{}, and
  Stable Audio Open. For \ours{}, the rapidly varying (high) latent half decodes to
  audio with a lower spectral centroid, matching Table~\ref{tab:latent_probe}.}
  \label{fig:latent_probe}
\end{figure}

\paragraph{Observation and interpretation.}
\ours{} ($\rho{=}2.36$) and \earvae{} ($\rho{=}1.82$) are inverted. LeVo~2, SAME-L,
and Stable Audio Open are not ($\rho<1$), and \ours{} has the largest inversion ratio.
Both \ours{} and \earvae{} use the perceptual training losses introduced by \earvae{},
but this comparison does not isolate their effect. Figure~\ref{fig:latent_probe}
visualizes the measured separation.

This result is correlational and does not establish a causal mechanism or
a quality advantage. Spectral centroid measures brightness, not rhythm or onset
structure. The models differ in architecture, latent configuration, sample rate,
objective, and compression, and partial latents are outside the decoder's training
distribution. A causal study requires matched architectures, loss ablations, multiple
checkpoints, and a larger track set.

\end{document}